\documentclass[%
 reprint,
superscriptaddress,
 amsmath,amssymb,
 aps,
prb,
]{revtex4-2}

\usepackage{graphicx}
\usepackage{dcolumn}
\usepackage{bm}
\usepackage{siunitx}
\usepackage[dvipsnames]{xcolor}

\begin{document}


\title{Electronic Impurity Scattering Induced Spin Accumulation in Metallic Thin Films}

\author{Ming-Hung Wu}
\email{mh.wu@bristol.ac.uk}%
\affiliation{H. H. Wills Physics Laboratory, University of Bristol, Bristol BS8 1TL, United Kingdom}
\author{Alexander Fabian}
\affiliation{Institute for Theoretical Physics, Justus Liebig University Giessen, Heinrich-Buff-Ring 16, 35392 Giessen, Germany}
\affiliation{Center for Materials Research (LaMa), Justus Liebig University Giessen, Heinrich-Buff-Ring 16, 35392 Giessen, Germany }
\author{Martin Gradhand}%
\affiliation{H. H. Wills Physics Laboratory, University of Bristol, Bristol BS8 1TL, United Kingdom}
\affiliation{Institute of Physics, Johannes Gutenberg University Mainz, 55099 Mainz, Germany}%

\date{\today}

\begin{abstract}
In order to explore the spin accumulation, evaluating the spin galvanic and spin Hall effect, we utilize the semi-classical Boltzmann equation based on input from the relativistic Korringa-Kohn-Rostoker Green's function method, within the density functional theory. We calculate the spin accumulation including multiple contributions, especially skew-scattering (scattering-in term) and compare this to three different approximations, which include the isotropic and anisotropic relaxation time approximation. For heavy metals, with strong intrinsic spin-orbit coupling, we find that almost all the effects are captured within the anisotropic relaxation time approximation.  On the other hand, in light metals the contributions from the vertex corrections (scattering-in term) are comparable to the induced effect in anisotropic relaxation time approximation. We put a particular focus on the influence of the atomic character of the substitutional impurities on the spin accumulation as well as the dependence on the impurity position. As impurities will break space inversion symmetry of the thin film, this will give rise to both symmetric and antisymmetric contributions to the spin accumulation. In general, we find the impurities at the surface generate the largest efficiency of charge-to-spin conversion in case of the spin accumulation. Comparing our results to existing experimental findings for Pt we find a good agreement.

\end{abstract}

\maketitle


\section{INTRODUCTION  \label{subsec:Introduction}}

The spin-orbit torque, a spin dynamics effect intimately connected to the spin-orbit coupling (SOC), has been widely investigated for its promising technological applications, ~\cite{Apalkov2016,Ando2015,Zhao2015,Zutic2004,RALPH2008}. For any practical application devices are built in stacks of thin films from multiple materials, which leads not only to spin currents but spin accumulation at the interface. The spin accumulation induced via Rashba spin-orbit coupling was already introduced by V.M.~Edelstein \cite{Edlestein1990, Bychkov1984}, and experimentally demonstrated by J.C.R.~S{\'a}nchez \textit{et al.} \cite{Sanchez2013}, partially making the connection to the spin Hall effect~ \cite{Smit1955,Hirsch1999,Kato2004,Sinova2004,Sinova2015,Nagaosa2010}.

In reality, it is impossible to grow solids, surfaces and interfaces for actual devices in perfect structures without any disorders. Thus besides the spin accumulation from the intrinsic SOC such as the Rashba-Edelstein effect and intrinsic SHE~\cite{Chang1995PRB,Chang1995PRL,Sundaram1999,Edlestein1990}, the extrinsic effects originating from spin-dependent scattering at disorder potentials simultaneously play a critical role in the physics of spin accumulation. While from experiments it is evident that impurities and disorder at the interfaces play an important role~\cite{Manchon2019,Zhu2019,Gueckstock2021} their effect on the spin accumulation has not been addressed explicitly in theoretical descriptions~\cite{Nikolic2005, Nikolic2006,Geranton2016,Fabian2021}.

This motivates us to investigate the spin accumulation including spin-dependent electron scattering at impurities, which will include contributions from the inversion symmetry breaking via the impurity potentials. For the spin Hall effect it has been shown previously that impurity doping may generate giant effects~\cite{Kimura2007,Gradhand2010_PRB020403,Kato2004}, even with relatively light elements~\cite{Gradhand2010PRL}. Moreover in thin films, the naturally imperfect structures at interfaces and surfaces, will inevitably contribute to spin-dependent scattering. These effects have been analysed experimentally~\cite{Kato2004,Zhu2019,Gueckstock2021,Avci2014} and theoretically~\cite{Herschbach2014,Saidaoui2016,Freimuth2014,Garello2013,Geranton2016} and were shown to exhibit possibly giant contributions to the overall effect. This implies the particular relevance of impurity position around the surface in these thin metallic films. 

In this article, we will use density functional theory (DFT) solved by the fully relativistic Korringa-Kohn-Rostoker (KKR) Green’s function method to investigate the current-induced spin accumulation properties of various materials and the effect of impurities. In Sec. \ref{sec:ComputationalMethods}, we briefly introduce the computational framework and various approximations for the description of the transport properties which are all based on the semi-classical Boltzmann equation~\cite{Gradhand2010_PRB020403,Gradhand2010_PRB245109,Geranton2016,Fabian2021} either including or ignoring the scattering-in term (vertex corrections). The structural details of the thin films will be discussed in detail. In Sec. \ref{sec:Results}, we present the results of the induced spin accumulation for thin metallic films of 9 monolayers (ML) within the distinct approximations in order to analyse the different contributions. To gain a better understanding of the influence of the atomic character on the effect we compare various combinations of host and impurity atoms. Making contact with experiment and for comparison among different alloys we introduce the normalized spin accumulation, a measure for the efficiency of charge-to-spin-accumulation conversion. Finally, we explain some of the findings in Pt based thin films in terms of the host density of states (DOS) as well as the impurity LDOS. A summary of our findings is presented in Sec. \ref{sec:summary}.

\section{Computational Methods and Details \label{sec:ComputationalMethods}}

All electronic structures, including the ideal and impurity systems, are calculated based on density functional theory (DFT) within the local density approximation solved by the relativistic Korringa-Kohn-Rostoker Green's function method \cite{KORRINGA1947,Kohn1954,Gradhand2009}.

\begin{figure}[t]
\includegraphics[width=0.46\textwidth]{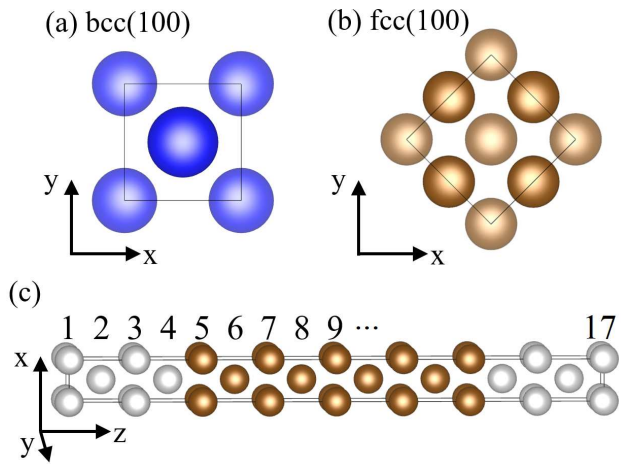}
\caption{\label{fig:schematic illustration}
(a), (b) The crystal structure and in-plane direction of bcc(100) and fcc(100). The transparency degree displays different layers. (c) Schematic of thin films fcc(100) structure. Atom 1 to 4 (and 14 to 17) are the vacuum spheres and 5 to 13 are the metallic atoms.
}
\end{figure}

For the description of the spin accumulation the semi-classical linearized Boltzmann equation \cite{Gradhand2010PRL} is solved to find the mean free path 
\begin{equation}
    {\boldsymbol \Lambda}^{n}(\mathbf{k})=\tau_{\mathbf{k}}^{n}\left[{\bf v}_{\mathbf{k}}^{n}+\sum_{\mathbf{k}^{\prime} n^{\prime}} P_{\mathbf{k}^{\prime} \mathbf{k}}^{n^{\prime} n} {\boldsymbol\Lambda}^{n^{\prime}}\left(\mathbf{k}^{\prime}\right)\right]\ \mathrm{,}
    \label{eq:Boltzmann equation}
\end{equation}
where $n$, \(\tau_\textbf{k}^n\) and \({\bf v}_\textbf{k}^n\) are band index, momentum relaxation time and Fermi velocity, respectively. The scattering rate \(P_{\textbf{k}'\textbf{k}}^{n'n}\) is calculated in the dilute limit from Fermi's golden rule, given by
\begin{equation}
    P_{\mathbf{k}^{\prime}\mathbf{k}}^{n^{\prime} n}=\frac{2 \pi}{\hbar} c_{0} N\left|T_{\mathbf{k }^{\prime}\mathbf{k}}^{n^{\prime}n}\right|^{2} \delta\left(E_{\mathbf{k}^{\prime}}^{n^{\prime}} - E_{\mathbf{k}}^{n}\right),
    \label{eq:scattering_rate}
\end{equation}
where \(c_{0} N\) is the number of impurities and \(T_{\mathbf{k}^{\prime}\mathbf{k}}^{n^{\prime}n}\) is the transition matrix describing the scattering of Bloch waves by the impurity potential. Thus all these quantities are obtained from the fully-relativistic electronic structure calculations of the host as well as the impurity system. For simplicity we define the charge currents in the $x$ direction (see Fig.~\ref{fig:schematic illustration}). Therefore, the charge conductivity is given by 
\begin{equation}
    \sigma_{xx}=\frac{e^{2}}{\hbar} \sum_{n} \frac{1}{(2 \pi)^{2}} \int_{E_{F}} \frac{d S_{n}}{\left|\mathbf{v}_{\mathbf{k}}^{n}\right|} v_{x, \mathbf{k}}^{n} \Lambda_{x}^{n}(\mathbf{k}),
    \label{eq:Charge conductivity}
\end{equation}
and the spin accumulation is expressed as the current-induced magnetization~\cite{Geranton2016}
\begin{equation}
    \chi^i_{yx}=\frac{e}{\hbar} \sum_{n} \frac{ \mu_{B} V}{d(2 \pi)^{2}} \int_{E_{F}} \frac{d S_{n}}{\left|\mathbf{v}_{\mathbf{k}}^{n}\right|} s_{y, \mathbf{k}}^{n,i} \Lambda_{x}^{n}(\mathbf{k}),
    \label{eq:normal magnetization}
\end{equation}
where $V$ is the volume of the cell, $d$ is the thickness of the film, and $s_{y, \mathbf{k}}^{n,i}$ is the spin expectation value in the $y$ direction. The index $i$ labels the atomic position in $z$-direction perpendicular to the thin film surface as introduced in Fig.~\ref{fig:schematic illustration}. $\chi^i_{yx}$ represents the magnetization along the $y$-direction induced by the charge current in $x$-direction as a function of the atomic layer index $i$. If Eq.~(\ref{eq:Boltzmann equation}) is solved and $\Lambda_{x}^{n}(\mathbf{k})$ is used in equation (\ref{eq:normal magnetization}) it implies that all scattering processes, including the scattering-in term, are fully taken into account. In order to gain further insight into the various mechanisms which are partially opposing each other, we will define 3 distinct approximations. First, we may evaluate the spin accumulation by dropping the scattering-in term $P_{\mathbf{k}^{\prime} \mathbf{k}}^{n^{\prime} n} {\boldsymbol\Lambda}^{n^{\prime}}\left(\mathbf{k}^{\prime}\right)$ in Eq.~(\ref{eq:Boltzmann equation}) leading to the anisotropic relaxation time approximation
\begin{equation}
\mathring{\Lambda}_{x}^{n}(\mathbf{k})=\tau_{\mathbf{k}}^{n} v_{x,\mathbf{k}}^{n}\ \text{.}
\end{equation}
The resulting spin accumulation is
\begin{equation}
    \mathring{\chi}^i_{yx}=\frac{e}{\hbar} \sum_{n} \frac{ \mu_{B} V}{d(2 \pi)^{2}} \int_{E_{F}} \frac{d S_{n}}{\left|\mathbf{v}_{\mathbf{k}}^{n}\right|} s_{y, \mathbf{k}}^{n,i} \mathring{\Lambda}_{x}^{n}(\mathbf{k})\ \text{.}
    \label{eq:ring magnetization}
\end{equation}
Second, we may define 
\begin{equation}
    \tilde{\chi}_{yx}^i=\frac{e}{\hbar} \sum_{n} \frac{ \mu_{B} V}{d(2 \pi)^{2}}  \int_{E_{F}} \frac{d S_{n}}{\left|\mathbf{v}_{\mathbf{k}}^{n}\right|} s_{y, \mathbf{k}}^{n,i} \left[ \Lambda_{x}^{n}(\mathbf{k})-\mathring{\Lambda}_{x}^{n}(\mathbf{k})\right]
    \label{eq:underline magnetization}
\end{equation}
in order to quantify the bare contributions to the spin accumulation arising from the scattering-in term. Third, addressing contributions from the clean system only, the isotropic relaxation time $\bar{\tau}$ can be used to define~\cite{Fabian2021}
\begin{equation}
    \bar{\chi}_{yx}^i=\frac{e}{\hbar} \sum_{n} \frac{\mu_{B} V \bar{\tau}}{d(2 \pi)^{2}} \int_{E_{F}} \frac{d S_{n}}{\left|\mathbf{v}_{\mathbf{k}}^{n}\right|} s_{y, \mathbf{k}}^{n,i} v_{x, \mathbf{k}}^{n}.
    \label{eq:tilde magnetization}
\end{equation}
Comparing, the full calculation to these three approximations will give us a deeper understanding of the underlying microscopic processes contributing to the spin accumulation in thin metallic films.

For practical applications the actual spin accumulation is only part of the relevant parameters to classify the efficiency of the spin-conversion mechanism in materials. To quantify this efficiency we introduce the normalized spin accumulation $\alpha^i_{yx}$, similar to the conventionally used spin Hall angle, as
\begin{equation}
    \alpha^i_{yx} = \frac{a^i_y}{j_x} = \frac{\chi^i_{yx}}{\sigma_{xx}},
\end{equation}
where $a^i_y$ and $j_x$ are the induced magnetic moments along the $y$ direction and the $x$-direction current density, respectively. They are defined via the electric field $E_x$ in the $x$ direction as $a_y = \chi_{yx} E_{x}$ and $j_x = \sigma_{xx} E_{x}$.

Finally, the model systems in our calculations are simple cubic crystals, fcc Cu(100), fcc Pt(100) and bcc U(100) thin films. All systems are built from 9 ML conserving the bulk inversion symmetry. The $x$-$y$ crystalline planes are defined in Figure 1 (a) and (b). Along the $z$ direction the vacuum layers are extended to infinity. The layer index $i$ is defined in Figure~1~(c) as used in the following results section. The angular momentum cutoff is $l_{max}=3$ and the number of k-points for the self-consistency of the clean system is $64 \times 64$. All the lattice constants are experimental values based on bulk crystals.

\section{Results \label{sec:Results}}

\begin{figure}[t]
\includegraphics[width=0.48\textwidth]{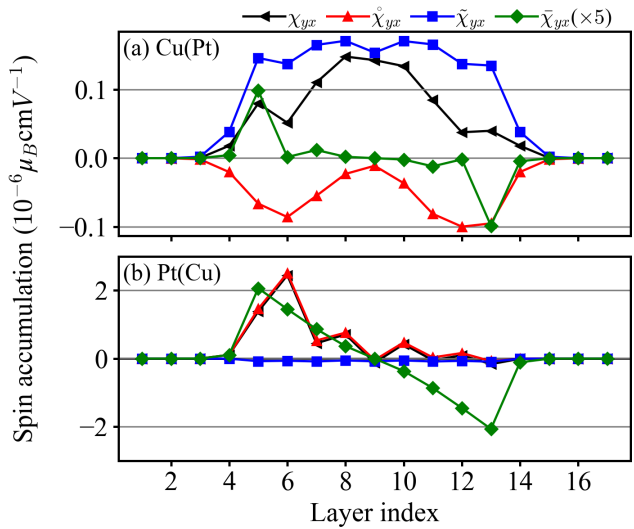}
\caption{\label{fig:PtCuembed}
The spin accumulation $\chi_{yx}$, $\mathring{\chi}_{yx}$, $\bar{\chi}_{yx}$ and $\tilde{\chi}_{yx}$ for a) Cu(Pt) and b) Pt(Cu) with the impurity placed in the surface layer $i=5$.
}
\end{figure}

In Fig.~(\ref{fig:PtCuembed}) the results are summarized for a Cu thin film with Pt impurities, Cu(Pt) as well as the inverted system with Cu impurities in a Pt thin film, Pt(Cu). The materials were chosen as Pt is known to exhibit a large intrinsic spin Hall conductivity and is one of the standard charge to spin as well as spin to charge conversion materials \cite{Sagasta2016,Liu2011_PRL,Ando2011_PRL}. In contrast, Cu is known to show a small intrinsic spin Hall angle~\cite{Sinova2015} but may be doped with heavy impurities to exhibit gigantic spin Hall angles induced by skew scattering (scattering-in term) \cite{Gradhand2010PRL}. For this comparison the impurity atom is placed in the surface layer ($i=5$) breaking the inversion symmetry of the thin films. Breaking this symmetry implies both symmetric as well as antisymmetric contributions will influence the overall spin accumulation.   

A number of conclusions can be drawn from the presented comparison. First of all the effect in Pt is an order of magnitude stronger than in the Cu thin film. Second, while for the Pt thin film the effect arises almost entirely within the anisotropic relaxation time approximation, $\mathring{\chi}_{yx}$, with vanishing contributions from the scattering-in term ($\tilde{\chi}_{yx}$), this is dramatically different for the Cu thin film. For Cu it turns out the contributions in anisotropic relaxation time approximation are opposing the effect induced by the scattering-in term while they are of similar order of magnitude. Third, in both cases the induced spin accumulation in isotropic relaxation time approximation $\bar{\chi}_{yx}$ is vanishingly small (scaled by factor 5 in Fig.~\ref{fig:PtCuembed}) and contributes marginally to the overall result. 

All together this leads to the fact that at the surface of the Cu thin film, for the atoms with the index 5 and 11, the spin accumulation is dominated by the extrinsic contributions as induced via the scattering-in term. This finding is in agreement with Ref.~\cite{Gueckstock2021} where it was suggested that the spin accumulation at the interfaces is dominated by the extrinsic skew-scattering mechanism. 

Furthermore, it is important to mention that in Cu the dominant contributions are almost symmetric with respect to the inversion center of the thin film whereas for Pt a strong antisymmetric component can be identified. In agreement with previous work~\cite{Fabian2021}, the induced accumulation in the isotropic relaxation time approximation is perfectly antisymmetric resulting in dominant contributions at the surfaces.

\begin{figure}[t]
\includegraphics[width=0.48\textwidth]{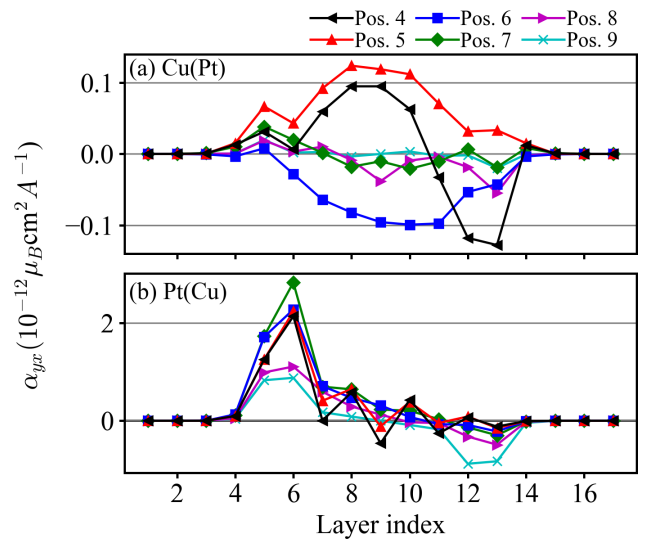}
\caption{\label{fig:normalisedSA}
Comparison of the normalised spin accumulation $\alpha_{yx}$ for the various impurity positions in the Cu(Pt) and Pt(Cu) thin film.
}
\end{figure}

Comparing both systems it appears natural to associate the different findings with the underlying electronic structure of the clean materials. While Pt shows strong spin-orbit coupling this is small for Cu. Resulting from this, it is the scattering at the impurity including the scattering-in term which dominates all effects in Cu. In contrast the strong spin-orbit coupling of clean Pt enables a strong spin accumulation with contributions almost exclusively arising within the anisotropic relaxation time approximation, excluding the scattering-in terms. Furthermore, the effect in Pt is much more restricted to the surface whereas for Cu the spin-accumulation is of similar strength across the thin film. This finding can be associated to the much longer spin-diffusion length of Cu in contrast to Pt, which will lead to a quick decay of the spin accumulation in a Pt thin film \cite{Sagasta2017,Kimura2005,Zhang2013,Gradhand2010_PRB245109}. 

\begin{table*}[t!]
\begin{ruledtabular}
\begin{tabular}{ccccccc}
Element&
Pos. 4&
Pos. 5&
Pos. 6&
Pos. 7&
Pos. 8&
Pos. 9\\
\hline
Sc & 1.31  & 0.64 & 0.37 & 0.36 & 0.25 & 0.33 \\
Ti & 0.74  & 0.61 & 0.35 & 0.34 & 0.24 & 0.31 \\
V  & 0.54  & 0.57 & 0.32 & 0.31 & 0.22 & 0.28 \\
Cr & 0.48  & 0.62 & 0.33 & 0.31 & 0.21 & 0.27 \\
Mn & 0.53  & 0.88 & 0.48 & 0.39 & 0.26 & 0.33 \\
Fe & 0.60  & 2.1  & 1.16 & 0.80 & 0.56 & 0.59 \\
Co & 0.78  & 9.11 & 4.06 & 2.70 & 1.93 & 1.91 \\
Ni & 1.65  & 4.17 & 1.67 & 1.40 & 1.11 & 1.43 \\
Cu & 17.32 & 1.09 & 0.60 & 0.52 & 0.40 & 0.51 \\
Zn & 5.57  & 0.82 & 0.47 & 0.43 & 0.32 & 0.41 \\
\end{tabular}
\end{ruledtabular}
\caption{\label{tab:sigmaxx}%
Charge conductivity of the Pt thin films doped with various impurities at different positions for the nonmagnetic systems. The unit is in $(\mu \Omega \textup{cm})^{-1}$.
}
\end{table*}

In a next step we consider the scenario where the impurities are placed at different positions across the thin film, ultimately preserving the structural inversion symmetry for the case of the impurity being placed at the central layer ($i=9$). The summary of the data is shown in Fig.~\ref{fig:normalisedSA}, varying the impurity position from the central atom ($i=9$) to the surface ($i=5$) up to the adatom position ($i=4$). In order to compare the different impurity positions we chose the normalized spin-accumulation as defined in Eq.~(\ref{eq:normal magnetization}). To be able to extract the bare spin accumulation the charge conductivities are summarized in Table~\ref{tab:sigmaxx} for Pt and in Table~\ref{tab:Cu_U_conductivty} for Cu.

As expected from the previous discussion, the effect in Cu is dramatically altered by the impurity position as the system is driven by extrinsic scattering. The induced spin accumulation even changes sign as the impurity is buried deeper in the thin film ($i=6$) and becomes more symmetric for the impurity at the adatom position. This dramatic effect of the impurity position on the overall spin accumulation was already discussed in experiments~\cite{Gueckstock2021}. The resulting spin accumulation turns perfectly antisymmetric as the impurity is placed at the central position ($i=9$) as the structural inversion symmetry is preserved in that case. The behaviour for the case of the Pt thin film is quite different as this system is driven by the intrinsic spin-orbit coupling of the Pt host. While similarly to Cu the overall magnitude of the effect drops as the impurity is buried deeper in the thin film the overall structure across the film does change less significantly with the sign of the spin accumulation staying constant at the surface with the layer index $i=5$ and $i=6$. Interestingly, it was found previously that the spin Hall angle for bulk systems of Cu(Pt) is larger than for Pt(Cu), with spin Hall angles of $27.0\times10^{-3}$ and $-5.2\times10^{-3}$, respectively~\cite{Gradhand2010PRL}. This is no longer true for the spin accumulation in thin films where the normalized spin accumulation of Cu(Pt) is significantly smaller than the effect in Pt(Cu). This highlights the importance of quantitative predictions in realistic geometries as they might be dramatically different from simplified descriptions in infinite bulk systems.

In order to get a more intuitive understanding of the induced spin accumulation for various impurities we calculated the normalised spin accumulation for the full $3d$ series from Sc to Zn placed inside a Pt thin film (see Tables~\ref{tab:sigmaxx}~and~\ref{tab:largest_NSA_of_Pt}). We find that all the spin accumulation profiles remain similar to the case of Cu impurities. In all cases the normalized spin accumulation is highest at the layer indices $i=5,6,7$ and it is dominated by the contributions in the anisotropic relaxation time approximation with minimal contributions from the scattering-in term. This is to be expected as the intrinsic spin-orbit coupling of the Pt host plays the dominant part for all impurities considered here.

While there is a large amount of literature on the theoretical description of the spin Hall conductivities in bulk systems~\cite{Sinova2004,Sundaram1999,Guo2008,Gradhand2010PRL,Gradhand2011,Fert2011,Seemann2015,Lowitzer2011,Zimmermann2014,Sagasta2016,Lowitzer2010,Freimuth2010,Geranton2016}, the studies on the spin accumulation are sparse. In Ref.~\cite{Geranton2016} a value for the normalized spin accumulation for Pt interfacing with L1$_{0}$-FePt was reported as up to $17.2\times10^{-13}\mu_{B}\textup{cm}^2\textup{A}^{-1}$. This is comparable to our maximum $\alpha_{yx}$ of Pt of the order of $\sim10^{-12}\mu_{B}\textup{cm}^2\textup{A}^{-1}$ for the various 3$d$ transition impurities. In comparison to experiment our results are compatible to the value of $\alpha_{yx} = 5 \times 10^{-12} (\mu_{B} \textup{cm}^{2} \textup{A}^{-1})$ for a film with thickness t $\geq$ 40 nm as measured in Ref.~\cite{Stamm2017}. The same group also reported measurements for a Pt(10 nm)/Cu(10 nm) bilayer system and found the magnetic moments of $a_y\approx 1.5\times10^{-6}\mu_B$ in Pt with an injected current density of $ 2.6\times 10^{6} \textup{A} \textup{cm}^{-2}$ ~\cite{Stamm2019}. This would result in a normalized spin accumulation of $\alpha_{yx}=5.7 \times 10^{-13}(\mu_{B} \textup{cm}^{2} \textup{A}^{-1})$ again comparable to our prediction.

For Cu, the experimental observations for the normalized spin accumulation vary between $6 \times 10^{-9}\mu_{B} \textup{cm}^{2} \textup{A}^{-1}$ and $3 \times 10^{-12} \mu_{B} \textup{cm}^{2} \textup{A}^{-1}$ depending on the experimental situation~\cite{Ding2020,Kukreja2015}. These are dramatically larger than our predictions of the order of $\sim10^{-13}\mu_{B} \textup{cm}^{2} \textup{A}^{-1}$. Partially, this discrepancy might be explained by the fact that in experiments the detected spin accumulation for Cu was measured at the interface with ferromagnetic materials. There, the proximity effect of magnetism could be considerable. Furthermore, the spin current in the experiments is generated by spin-pumping and spin-injection methods measuring the inverse effect of spin-to-charge conversion. In our calculations we consider the direct charge-to-spin conversion with no charge currents perpendicular to the surface and in the absence of any additional external voltages.

As was discussed already, the longitudinal charge conductivity $\sigma_{xx}$ is critical for the efficiency of the charge-to-spin conversion. In addition, a large charge conductivity would make any effect unobservable in real systems as the transport properties would be determined by residual resistivities unrelated to the specific impurity doping we are considering here. All longitudinal conductivities for the Pt based system are summarized in Table \ref{tab:sigmaxx}. The conventionally magnetic elements, such as Mn, Fe, Co and Ni, are enforced to be nonmagnetic to simplify the situation focusing purely on the effect induced via spin-orbit coupling disconnected from magnetism. The results for the magnetic cases including charge conductivity and spin accumulation are summarized and discussed in the Appendix~\ref{sec:Appendix}. In general, $\sigma_{xx}$ is largest for the surface position as the impurity only weakly interacts with the thin film electronic structure. As the impurities are buried deeper into the thin film the conductivities drop significantly only to get enhanced at the central position again ($i=9$) which is a finite size effect of these perfect thin films. In a simplified picture of a particle in a box every second wave function would show a node in the central position as such avoiding the scattering by the impurity. This leads to the slight enhancement of the conductivity for this central position.

\begin{figure}[t]
\includegraphics[width=0.48\textwidth]{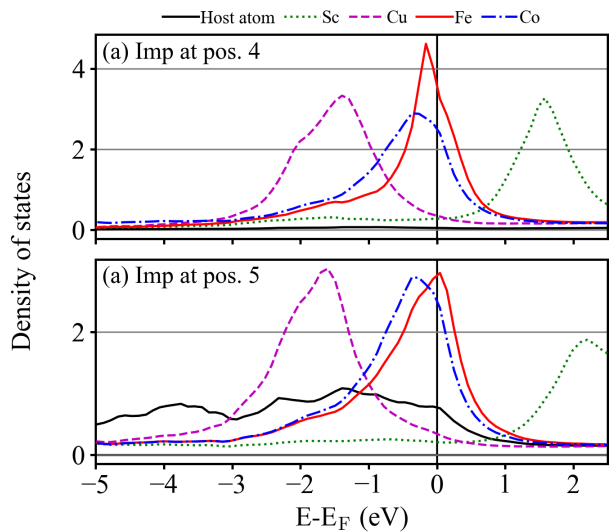}
\caption{\label{fig:PtDOS}
Comparison of density of states for the Sc, Cr, nonmagnetic Fe and Cu as the adatom ($i=4$) or as surface atom ($i=5$).
}
\end{figure}

To better understand the mechanism leading to the largely different conductivities for different impurities, we show the local density of states (LDOS) of the host atom and the doped impurities, Sc, Fe, Co and Cu at layer index $i=4$ and $i=5$ for the nonmagnetic systems in Fig.~\ref{fig:PtDOS}. For the adatom the host LDOS (an empty sphere), shows almost vanishing contributions, whereas Co and Fe exhibit strong peaks in the LDOS at the Fermi energy implying reasonably strong scattering. In contrast, the Cu and Sc impurity show a low LDOS at the Fermi energy leading to weak scattering. This intuitively explains the considerably lower charge conductivity for the Fe and Co impurity in comparison to Sc and Cu.  

In slight contrast the host LDOS at layer index $i=5$, the surface atom, shows large contributions of predominantly $d$ character. In this case the $d$ electron peaks of Co and Fe lead to weaker scattering than the $s$ and $p$ character LDOS of Sc and Cu at Fermi level. This in turn leads to the opposite effect to the adatom situation where now the conductivities are higher for Fe and Co but are suppressed for Sc and Cu.

\begin{figure}[t]
\includegraphics[width=0.48\textwidth]{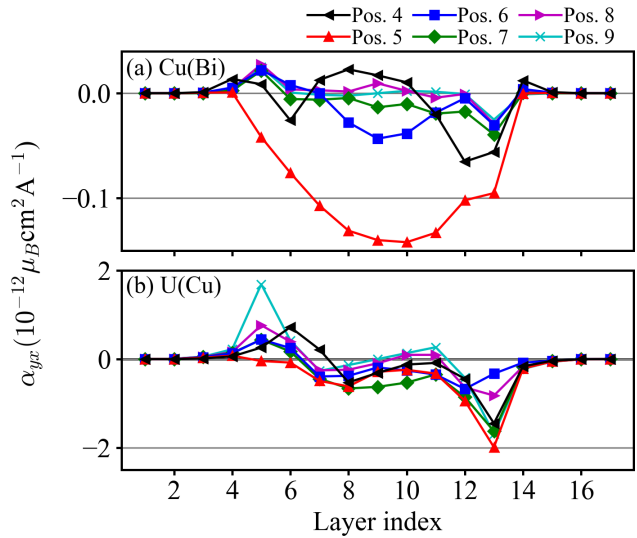}
\caption{\label{fig:CuBi_UCu_normalisedSA}
Comparison of the normalised spin accumulation $\alpha_{yx}$ for the various impurity positions in the Cu(Bi) and U(Cu) thin films.
}
\end{figure}

So far we focused on the $d$-electron host systems with strong spin-orbit coupling such as Pt. This system is overall weakly affected by the impurity character whereas for the small spin-orbit coupling Cu system previous calculations have indicated that the spin-dependent transport is significantly altered by the choice of the impurity atom ~\cite{Gradhand2010PRL,Fert2011}.
One of the largest spin Hall angles was found for Cu(Bi)~\cite{Gradhand2010_PRB245109,Niimi2012}. Similarly, we may try to enhance the effect by choosing a system with even stronger spin-orbit coupling, uranium, the heaviest naturally occurring metal ~\cite{Yoo1998,Barrett1963,Springell2008,Wilson1949}. In Fig.~\ref{fig:CuBi_UCu_normalisedSA} (a), the layer dependence of the normalized spin accumulation is shown for Cu(Bi), which turns out to be comparable to the Cu(Pt) system or even smaller with notably a different sign for the effect. While the spin Hall conductivity was found to be dramatically enhanced in a Cu(Bi) bulk system, both theoretically~\cite{Gradhand2010PRL,Fedorov2013} and experimentally~\cite{Niimi2012}, it turns out that the spin accumulation is rather suppressed for the Bi doped Cu thin film.

In Fig.~\ref{fig:CuBi_UCu_normalisedSA} (b), we summarize the result for the bcc uranium thin films doped with Cu hoping to exploit the large spin-orbit coupling of U. It turns out that for almost all scenarios the resulting normalized spin accumulation is comparable to the Pt based system. Furthermore, an impurity placed at the surface position induces the largest spin accumulation at the opposing side of the thin film which appears counter intuitive and contrasts with the results for the Pt thin films. 

The resulting charge conductivities of doping a series of impurities into the Cu and U thin films are summarized in Table~\ref{tab:Cu_U_conductivty}. In the Cu films the magnitude for Pt impurities is an order of magnitude larger than for the Bi despite inducing a similar normalized spin accumulation. For all scenarios of U thin films they are of similar order of magnitude with the exception of the impurities at the adatom position where we find a wide variation. Table \ref{tab:Cu_U_NSA} shows the resulting normalized spin accumulation for all cases. Interestingly, for the U thin films, we find that Cu induces a similar normalized spin accumulation to the other impurities. This strong Cu-induced accumulation differs notably from Ref.~\cite{Wu2020} that demonstrated a vanishing spin Hall angle for Cu doped bcc bulk U. The reason for this discrepancy is the restriction to the skew-scattering mechanism in Ref.~\cite{Wu2020}, while here we incorporate contributions arising from the spatial inversion symmetry breaking induced by the impurity scattering. Such contributions are not present in bulk systems where substitutional impurities do not break spatial inversion symmetry. However, in thin films as considered here any impurity not placed at the central layer would break the symmetry, inducing spin accumulations in addition to the skew-scattering contribution.

\begin{table}[t]
\begin{ruledtabular}
\begin{tabular}{ccccccc}
\textrm{Element}&
\textrm{Pos. 4}&
\textrm{Pos. 5}&
\textrm{Pos. 6}&
\textrm{Pos. 7}&
\textrm{Pos. 8}&
\textrm{Pos. 9}\\
\colrule
Sc      & 1.28 & 2.30 & 2.02 & 2.77 & 1.03 & 1.07 \\
Ti      & 1.37 & 2.31 & 2.03 & 2.66 & 1.00 & 1.08 \\
V       & 1.45 & 2.33 & 2.11 & 2.54 & 0.97 & 1.06 \\
Cr      & 1.49 & 2.40 & 2.23 & 2.43 & 0.96 & 1.05 \\
Mn      & 1.37 & 2.64 & 2.33 & 2.31 & 0.97 & 0.99 \\
Fe      & 1.23 & 3.07 & 2.32 & 2.19 & 0.97 & 0.86 \\
Co      & 1.14 & 3.67 & 2.68 & 2.03 & 0.89 & 0.86 \\
Ni      & 1.15 & 1.99 & 2.36 & 2.95 & 1.12 & 0.87 \\
Cu      & 2.14 & 2.24 & 2.28 & 2.83 & 1.11 & 0.88 \\
Zn      & 1.34 & 2.30 & 2.14 & 2.83 & 1.07 & 0.96
\end{tabular}
\end{ruledtabular}
\caption{\label{tab:largest_NSA_of_Pt}%
The largest value of normalized spin accumulation $\alpha_{yx}$ of the Pt thin film doped with various impurities at different positions. For almost all systems, these values would be taken at the layer index $i=6$. The unit is in $10^{-12} \mu_B \textrm{cm} A^{-1}$.
}
\end{table}

\begin{table}[b]
\begin{ruledtabular}
\begin{tabular}{ccccccc}
Element&
Pos. 4&
Pos. 5&
Pos. 6&
Pos. 7&
Pos. 8&
Pos. 9\\
\colrule
Cu(Pt)        & 1.054 & 1.199 & 0.504 & 0.261 & 0.276 & 0.303 \\
Cu(Bi)        & 0.211 & 0.079 & 0.045 & 0.038 & 0.037 & 0.043 \\
              &       &       &       &       &       &       \\
U(Li)         & 0.870 & 0.108 & 0.075 & 0.064 & 0.062 & 0.055 \\
U(B)          & 0.357 & 0.089 & 0.076 & 0.058 & 0.056 & 0.053 \\
U(Al)         & 0.291 & 0.098 & 0.083 & 0.064 & 0.060 & 0.058 \\
U(nonmag. Fe) & 0.460 & 0.107 & 0.083 & 0.070 & 0.067 & 0.062 \\
U(Cu)         & 0.726 & 0.097 & 0.128 & 0.094 & 0.090 & 0.093 \\
\end{tabular}
\end{ruledtabular}
\caption{\label{tab:Cu_U_conductivty}%
Charge conductivity of the Cu and U thin film doped with Pt and Bi, and Li, B, nonmag. Fe and Cu, respectively. The unit is in $(\mu \Omega \textup{cm})^{-1}$.
}
\end{table}

\begin{table}[h]
\begin{ruledtabular}
\begin{tabular}{ccccccc}
Element&
Pos. 4&
Pos. 5&
Pos. 6&
Pos. 7&
Pos. 8&
Pos. 9\\
\colrule
Cu(Pt)       & 0.128 & 0.124 & 0.099 & 0.038 & 0.055 & 0.020 \\
Cu(Bi)       & 0.065 & 0.142 & 0.043 & 0.039 & 0.029 & 0.025 \\
              &       &       &       &       &       &       \\
U(Li)         & 1.25 & 1.92 & 0.70  & 1.18 & 0.76  & 0.87  \\
U(B)          & 0.85 & 1.66 & 0.68  & 1.13 & 0.93  & 0.87  \\
U(Al)         & 1.36 & 1.70 & 0.72  & 1.09 & 0.90  & 0.86  \\
U(nonmag. Fe) & 0.75 & 1.70 & 0.71 & 1.47 & 0.88  & 1.05  \\
U(Cu)         & 1.45 & 1.99 & 0.67 & 1.63 & 0.82 & 1.69 \\
\end{tabular}
\end{ruledtabular}
\caption{\label{tab:Cu_U_NSA}%
The largest magnitude of normalized spin accumulation $\alpha_{yx}$ of the Cu and U thin film doped with Pt and Bi, and Li, B, nonmag. Fe and Cu, respectively. The unit is in $10^{-12} \mu_B \textrm{cm} A^{-1}$.
}
\end{table}

\section{Summary  \label{sec:summary}}

In conclusion, we have utilized a DFT based Korringa-Kohn-Rostoker Green's function method to analyse the current-induced spin accumulation of impurity doped thin films, where the transport is described within a semi-classical Boltzmann formalism. The comparison of different approximations demonstrates that the spin accumulation for a Cu host, with weak spin-orbit coupling, is equally driven by the scattering-in term $\tilde{\chi}^i_{yx}$ as well as forward scattering processes covered in the anisotropic relaxation time approximation. However, both contributions are of opposite sign and partially cancel each other. The spin accumulation of Cu(Pt) is almost asymmetric under spatial inversion of the thin film. On the other hand, the spin accumulation of Pt(Cu) is almost entirely covered by the anisotropic relaxation time approximation $\mathring{\chi}^i_{yx}$ where a strong antisymmetric component exists. This leads to a vanishing spin accumulation at the opposite side, relative to the impurity position, of the thin film. For the isotropic relaxation time approximation, the induced spin accumulation in both cases is vanishingly small with symmetric profiles over the thin films. This highlights the importance of spin-dependent scattering processes in quantifying the spin accumulation in realistic materials. Moreover, the strong intrinsic SOC of Pt provides dramatically larger spin accumulation than a Cu host.

In addition, we analyse the normalized spin accumulation, the efficiency of charge-to-spin-accumulation conversion, in the Cu(Pt) and Pt(Cu) cases where the impurities are placed at different positions. While we found the effect of Cu(Pt) to be significantly dependent on the impurity position with a changing magnitude and sign. By doping the full 3$d$ series of impurities into the Pt thin film, we find that the all values and profiles of normalized spin accumulation remain similar, which highlights the importance of strong intrinsic spin-orbit coupling of the host. Furthermore, our results of normalised spin accumulation for the Pt thin film is in good agreement to experimental observations~\cite{Stamm2019,Stamm2017}, while our results of Cu thin film is at a first glance incompatible to experimental findings~\cite{Ding2020,Kukreja2015}. This discrepancy can be assigned to the fact that in experiments ferromagnetic materials at the interface to Cu might alter the actual spin accumulation significantly which remains an open topic for further investigations.

The spin accumulation for the heaviest naturally occurring metal, uranium, highlights a peculiar result where the spin accumulation is highest at the opposite side of the thin film relative to the impurity position. This is in stark contrast to our finding in Pt. While the generally more localized spin accumulation in heavy metals such as Pt and U can be associated to the reduced spin diffusion length in comparison to Cu, this result for U is not intuitively obvious and highlights the importance of actual material specific calculations in realistic materials to cover the complexity of all effects.

\begin{acknowledgments}
This work was sponsored by EPSRC international studentship EP/N509619/1. The authors are grateful to Michael Czerner and Christian Heiliger of the Institute for Theoretical Physics at the Justus Liebig University Giessen for technical support. M.G. thanks the visiting professorship program of the Centre for Dynamics and Topology ad Johannes Gutenberg-University Mainz.
\end{acknowledgments}

\appendix

\section{Charge conductivity and normalized spin accumulation induced by magnetic impurities}\label{sec:Appendix}

 As the magnetism is taken into account, the conventionally magnetic atoms such as Fe, Co and Ni as well as Mn develop a magnetic moment across all positions in the Pt thin films. For V the magnetism occurs only at the surface. All these magnetic atoms induce moments in the surrounding Pt coupling antiferromagnetically. For the transport calculations the independent moments of each impurity in the dilute alloy are assumed to point in the same direction, perpendicular to the thin films surface. 

Table~\ref{tab:mag_sigmaxx} summarizes the results for the charge conductivity for all magnetic impurities. In comparison to the nonmagnetic results in Table~\ref{tab:sigmaxx}, the conductivities are enhanced when the magnetic impurities are doped at layer index $i=4$ as adatoms. For Fe, Co and Ni, the conductivities are reduced as the impurities are buried deeper in the thin film while for Mn the conductivity becomes larger.

\begin{table}[h!]
\begin{ruledtabular}
\begin{tabular}{ccccccc}
\textrm{Element}&
\textrm{Pos. 4}&
\textrm{Pos. 5}&
\textrm{Pos. 6}&
\textrm{Pos. 7}&
\textrm{Pos. 8}&
\textrm{Pos. 9}\\
\colrule
V  & 0.80   & 0.60   &        &        &        &        \\
Mn & 2.82   & 0.90   & 0.54   & 0.52   & 0.38   & 0.48   \\
Fe & 1.12   & 0.74   & 0.40   & 0.39   & 0.28   & 0.36   \\
Co & 0.95   & 0.88   & 0.46   & 0.43   & 0.29   & 0.38   \\
Ni &        & 3.32   & 1.36   & 1.19   & 0.98   & 1.27   \\
\end{tabular}
\end{ruledtabular}
\caption{\label{tab:mag_sigmaxx}%
Charge conductivity $\sigma_{xx}$ of the Pt thin film doped with magnetic impurities at different positions in units $(\mu \Omega \textup{cm})^{-1}$.
}
\end{table}

The maximum normalized spin accumulations are summarized in Table~\ref{tab:mag_NSA}. For almost all cases the maximum is taken at the layer $i=6$. The results are comparable to the nonmagnetic case except for Fe and Co at the surface position ($i=5$), which are slightly reduced.

\begin{table}[h!]
\begin{ruledtabular}
\begin{tabular}{ccccccc}
\textrm{Element}&
\textrm{Pos. 4}&
\textrm{Pos. 5}&
\textrm{Pos. 6}&
\textrm{Pos. 7}&
\textrm{Pos. 8}&
\textrm{Pos. 9}\\
\colrule
V  & 1.42   & 2.38   &        &        &        &        \\
Mn & 1.40   & 2.36   & 2.18   & 2.82   & 1.05   & 1.00   \\
Fe & 1.44   & 2.43   & 2.17   & 2.70   & 1.02   & 1.00   \\
Co & 1.51   & 2.50   & 2.29   & 2.55   & 1.00   & 0.95   \\
Ni &        & 2.16   & 2.38   & 2.85   & 1.11   & 0.87   \\
\end{tabular}
\end{ruledtabular}
\caption{\label{tab:mag_NSA}%
The largest value of normalized spin accumulation $\alpha_{yx}$ of the Pt thin film doped with magnetic impurities at different positions. The unit is in $10^{-12} \mu_B \textrm{cm} A^{-1}$.
}
\end{table}




\begin{thebibliography}{61}%
\makeatletter
\providecommand \@ifxundefined [1]{%
 \@ifx{#1\undefined}
}%
\providecommand \@ifnum [1]{%
 \ifnum #1\expandafter \@firstoftwo
 \else \expandafter \@secondoftwo
 \fi
}%
\providecommand \@ifx [1]{%
 \ifx #1\expandafter \@firstoftwo
 \else \expandafter \@secondoftwo
 \fi
}%
\providecommand \natexlab [1]{#1}%
\providecommand \enquote  [1]{``#1''}%
\providecommand \bibnamefont  [1]{#1}%
\providecommand \bibfnamefont [1]{#1}%
\providecommand \citenamefont [1]{#1}%
\providecommand \href@noop [0]{\@secondoftwo}%
\providecommand \href [0]{\begingroup \@sanitize@url \@href}%
\providecommand \@href[1]{\@@startlink{#1}\@@href}%
\providecommand \@@href[1]{\endgroup#1\@@endlink}%
\providecommand \@sanitize@url [0]{\catcode `\\12\catcode `\$12\catcode
  `\&12\catcode `\#12\catcode `\^12\catcode `\_12\catcode `\%12\relax}%
\providecommand \@@startlink[1]{}%
\providecommand \@@endlink[0]{}%
\providecommand \url  [0]{\begingroup\@sanitize@url \@url }%
\providecommand \@url [1]{\endgroup\@href {#1}{\urlprefix }}%
\providecommand \urlprefix  [0]{URL }%
\providecommand \Eprint [0]{\href }%
\providecommand \doibase [0]{https://doi.org/}%
\providecommand \selectlanguage [0]{\@gobble}%
\providecommand \bibinfo  [0]{\@secondoftwo}%
\providecommand \bibfield  [0]{\@secondoftwo}%
\providecommand \translation [1]{[#1]}%
\providecommand \BibitemOpen [0]{}%
\providecommand \bibitemStop [0]{}%
\providecommand \bibitemNoStop [0]{.\EOS\space}%
\providecommand \EOS [0]{\spacefactor3000\relax}%
\providecommand \BibitemShut  [1]{\csname bibitem#1\endcsname}%
\let\auto@bib@innerbib\@empty
\bibitem [{\citenamefont {Apalkov}\ \emph {et~al.}(2016)\citenamefont
  {Apalkov}, \citenamefont {Dieny},\ and\ \citenamefont
  {Slaughter}}]{Apalkov2016}%
  \BibitemOpen
  \bibfield  {author} {\bibinfo {author} {\bibfnamefont {D.}~\bibnamefont
  {Apalkov}}, \bibinfo {author} {\bibfnamefont {B.}~\bibnamefont {Dieny}},\
  and\ \bibinfo {author} {\bibfnamefont {J.~M.}\ \bibnamefont {Slaughter}},\
  }\href@noop {} {\bibfield  {journal} {\bibinfo  {journal} {Proc. IEEE}\
  }\textbf {\bibinfo {volume} {104}},\ \bibinfo {pages} {1796} (\bibinfo {year}
  {2016})}\BibitemShut {NoStop}%
\bibitem [{\citenamefont {Ando}(2015)}]{Ando2015}%
  \BibitemOpen
  \bibfield  {author} {\bibinfo {author} {\bibfnamefont {Y.}~\bibnamefont
  {Ando}},\ }\href@noop {} {\bibfield  {journal} {\bibinfo  {journal} {Jpn. J.
  Appl. Phys.}\ }\textbf {\bibinfo {volume} {54}},\ \bibinfo {pages} {070101}
  (\bibinfo {year} {2015})}\BibitemShut {NoStop}%
\bibitem [{\citenamefont {Zhao}(2015)}]{Zhao2015}%
  \BibitemOpen
  \bibfield  {author} {\bibinfo {author} {\bibfnamefont {G.}~\bibnamefont
  {Zhao}, \bibfnamefont {W.~Prenat}},\ }\href@noop {} {\emph {\bibinfo {title}
  {Spintronics-Based Computing}}}\ (\bibinfo  {publisher} {Springer},\ \bibinfo
  {address} {Berlin},\ \bibinfo {year} {2015})\BibitemShut {NoStop}%
\bibitem [{\citenamefont {\ifmmode \check{Z}\else
  \v{Z}\fi{}uti\ifmmode~\acute{c}\else \'{c}\fi{}}\ \emph
  {et~al.}(2004)\citenamefont {\ifmmode \check{Z}\else
  \v{Z}\fi{}uti\ifmmode~\acute{c}\else \'{c}\fi{}}, \citenamefont {Fabian},\
  and\ \citenamefont {Das~Sarma}}]{Zutic2004}%
  \BibitemOpen
  \bibfield  {author} {\bibinfo {author} {\bibfnamefont {I.}~\bibnamefont
  {\ifmmode \check{Z}\else \v{Z}\fi{}uti\ifmmode~\acute{c}\else \'{c}\fi{}}},
  \bibinfo {author} {\bibfnamefont {J.}~\bibnamefont {Fabian}},\ and\ \bibinfo
  {author} {\bibfnamefont {S.}~\bibnamefont {Das~Sarma}},\ }\href
  {https://doi.org/10.1103/RevModPhys.76.323} {\bibfield  {journal} {\bibinfo
  {journal} {Rev. Mod. Phys.}\ }\textbf {\bibinfo {volume} {76}},\ \bibinfo
  {pages} {323} (\bibinfo {year} {2004})}\BibitemShut {NoStop}%
\bibitem [{\citenamefont {Ralph}\ and\ \citenamefont
  {Stiles}(2008)}]{RALPH2008}%
  \BibitemOpen
  \bibfield  {author} {\bibinfo {author} {\bibfnamefont {D.}~\bibnamefont
  {Ralph}}\ and\ \bibinfo {author} {\bibfnamefont {M.}~\bibnamefont {Stiles}},\
  }\href {https://doi.org/https://doi.org/10.1016/j.jmmm.2007.12.019}
  {\bibfield  {journal} {\bibinfo  {journal} {J. Magn. Magn. Mater.}\ }\textbf
  {\bibinfo {volume} {320}},\ \bibinfo {pages} {1190} (\bibinfo {year}
  {2008})}\BibitemShut {NoStop}%
\bibitem [{\citenamefont {Edelstein}(1990)}]{Edlestein1990}%
  \BibitemOpen
  \bibfield  {author} {\bibinfo {author} {\bibfnamefont {V.}~\bibnamefont
  {Edelstein}},\ }\href
  {https://doi.org/https://doi.org/10.1016/0038-1098(90)90963-C} {\bibfield
  {journal} {\bibinfo  {journal} {Solid State Commun.}\ }\textbf {\bibinfo
  {volume} {73}},\ \bibinfo {pages} {233} (\bibinfo {year} {1990})}\BibitemShut
  {NoStop}%
\bibitem [{\citenamefont {Bychkov}\ and\ \citenamefont
  {Rashba}(1984)}]{Bychkov1984}%
  \BibitemOpen
  \bibfield  {author} {\bibinfo {author} {\bibfnamefont {Y.~A.}\ \bibnamefont
  {Bychkov}}\ and\ \bibinfo {author} {\bibfnamefont {E.~I.}\ \bibnamefont
  {Rashba}},\ }\href {https://doi.org/10.1088/0022-3719/17/33/015} {\bibfield
  {journal} {\bibinfo  {journal} {J. Phys. C: Solid State Phys.}\ }\textbf
  {\bibinfo {volume} {17}},\ \bibinfo {pages} {6039} (\bibinfo {year}
  {1984})}\BibitemShut {NoStop}%
\bibitem [{\citenamefont {S{\'a}nchez}\ \emph {et~al.}(2013)\citenamefont
  {S{\'a}nchez}, \citenamefont {Vila}, \citenamefont {Desfonds}, \citenamefont
  {Gambarelli}, \citenamefont {Attan{\'e}}, \citenamefont {De~Teresa},
  \citenamefont {Mag{\'e}n},\ and\ \citenamefont {Fert}}]{Sanchez2013}%
  \BibitemOpen
  \bibfield  {author} {\bibinfo {author} {\bibfnamefont {J.~C.~R.}\
  \bibnamefont {S{\'a}nchez}}, \bibinfo {author} {\bibfnamefont
  {L.}~\bibnamefont {Vila}}, \bibinfo {author} {\bibfnamefont {G.}~\bibnamefont
  {Desfonds}}, \bibinfo {author} {\bibfnamefont {S.}~\bibnamefont
  {Gambarelli}}, \bibinfo {author} {\bibfnamefont {J.~P.}\ \bibnamefont
  {Attan{\'e}}}, \bibinfo {author} {\bibfnamefont {J.~M.}\ \bibnamefont
  {De~Teresa}}, \bibinfo {author} {\bibfnamefont {C.}~\bibnamefont
  {Mag{\'e}n}},\ and\ \bibinfo {author} {\bibfnamefont {A.}~\bibnamefont
  {Fert}},\ }\href {https://doi.org/10.1038/ncomms3944} {\bibfield  {journal}
  {\bibinfo  {journal} {Nat. Commun.}\ }\textbf {\bibinfo {volume} {4}},\
  \bibinfo {pages} {2944} (\bibinfo {year} {2013})}\BibitemShut {NoStop}%
\bibitem [{\citenamefont {Smit}(1955)}]{Smit1955}%
  \BibitemOpen
  \bibfield  {author} {\bibinfo {author} {\bibfnamefont {J.}~\bibnamefont
  {Smit}},\ }\href
  {https://doi.org/https://doi.org/10.1016/S0031-8914(55)92596-9} {\bibfield
  {journal} {\bibinfo  {journal} {Physica}\ }\textbf {\bibinfo {volume} {21}},\
  \bibinfo {pages} {877} (\bibinfo {year} {1955})}\BibitemShut {NoStop}%
\bibitem [{\citenamefont {Hirsch}(1999)}]{Hirsch1999}%
  \BibitemOpen
  \bibfield  {author} {\bibinfo {author} {\bibfnamefont {J.~E.}\ \bibnamefont
  {Hirsch}},\ }\href {https://doi.org/10.1103/PhysRevLett.83.1834} {\bibfield
  {journal} {\bibinfo  {journal} {Phys. Rev. Lett.}\ }\textbf {\bibinfo
  {volume} {83}},\ \bibinfo {pages} {1834} (\bibinfo {year}
  {1999})}\BibitemShut {NoStop}%
\bibitem [{\citenamefont {Kato}\ \emph {et~al.}(2004)\citenamefont {Kato},
  \citenamefont {Myers}, \citenamefont {Gossard},\ and\ \citenamefont
  {Awschalom}}]{Kato2004}%
  \BibitemOpen
  \bibfield  {author} {\bibinfo {author} {\bibfnamefont {Y.~K.}\ \bibnamefont
  {Kato}}, \bibinfo {author} {\bibfnamefont {R.~C.}\ \bibnamefont {Myers}},
  \bibinfo {author} {\bibfnamefont {A.~C.}\ \bibnamefont {Gossard}},\ and\
  \bibinfo {author} {\bibfnamefont {D.~D.}\ \bibnamefont {Awschalom}},\ }\href
  {https://doi.org/10.1126/science.1105514} {\bibfield  {journal} {\bibinfo
  {journal} {Science}\ }\textbf {\bibinfo {volume} {306}},\ \bibinfo {pages}
  {1910} (\bibinfo {year} {2004})}\BibitemShut {NoStop}%
\bibitem [{\citenamefont {Sinova}\ \emph {et~al.}(2004)\citenamefont {Sinova},
  \citenamefont {Culcer}, \citenamefont {Niu}, \citenamefont {Sinitsyn},
  \citenamefont {Jungwirth},\ and\ \citenamefont {MacDonald}}]{Sinova2004}%
  \BibitemOpen
  \bibfield  {author} {\bibinfo {author} {\bibfnamefont {J.}~\bibnamefont
  {Sinova}}, \bibinfo {author} {\bibfnamefont {D.}~\bibnamefont {Culcer}},
  \bibinfo {author} {\bibfnamefont {Q.}~\bibnamefont {Niu}}, \bibinfo {author}
  {\bibfnamefont {N.~A.}\ \bibnamefont {Sinitsyn}}, \bibinfo {author}
  {\bibfnamefont {T.}~\bibnamefont {Jungwirth}},\ and\ \bibinfo {author}
  {\bibfnamefont {A.~H.}\ \bibnamefont {MacDonald}},\ }\href
  {https://doi.org/10.1103/PhysRevLett.92.126603} {\bibfield  {journal}
  {\bibinfo  {journal} {Phys. Rev. Lett.}\ }\textbf {\bibinfo {volume} {92}},\
  \bibinfo {pages} {126603} (\bibinfo {year} {2004})}\BibitemShut {NoStop}%
\bibitem [{\citenamefont {Sinova}\ \emph {et~al.}(2015)\citenamefont {Sinova},
  \citenamefont {Valenzuela}, \citenamefont {Wunderlich}, \citenamefont
  {Back},\ and\ \citenamefont {Jungwirth}}]{Sinova2015}%
  \BibitemOpen
  \bibfield  {author} {\bibinfo {author} {\bibfnamefont {J.}~\bibnamefont
  {Sinova}}, \bibinfo {author} {\bibfnamefont {S.~O.}\ \bibnamefont
  {Valenzuela}}, \bibinfo {author} {\bibfnamefont {J.}~\bibnamefont
  {Wunderlich}}, \bibinfo {author} {\bibfnamefont {C.~H.}\ \bibnamefont
  {Back}},\ and\ \bibinfo {author} {\bibfnamefont {T.}~\bibnamefont
  {Jungwirth}},\ }\href {https://doi.org/10.1103/RevModPhys.87.1213} {\bibfield
   {journal} {\bibinfo  {journal} {Rev. Mod. Phys.}\ }\textbf {\bibinfo
  {volume} {87}},\ \bibinfo {pages} {1213} (\bibinfo {year}
  {2015})}\BibitemShut {NoStop}%
\bibitem [{\citenamefont {Nagaosa}\ \emph {et~al.}(2010)\citenamefont
  {Nagaosa}, \citenamefont {Sinova}, \citenamefont {Onoda}, \citenamefont
  {MacDonald},\ and\ \citenamefont {Ong}}]{Nagaosa2010}%
  \BibitemOpen
  \bibfield  {author} {\bibinfo {author} {\bibfnamefont {N.}~\bibnamefont
  {Nagaosa}}, \bibinfo {author} {\bibfnamefont {J.}~\bibnamefont {Sinova}},
  \bibinfo {author} {\bibfnamefont {S.}~\bibnamefont {Onoda}}, \bibinfo
  {author} {\bibfnamefont {A.~H.}\ \bibnamefont {MacDonald}},\ and\ \bibinfo
  {author} {\bibfnamefont {N.~P.}\ \bibnamefont {Ong}},\ }\href
  {https://doi.org/10.1103/RevModPhys.82.1539} {\bibfield  {journal} {\bibinfo
  {journal} {Rev. Mod. Phys.}\ }\textbf {\bibinfo {volume} {82}},\ \bibinfo
  {pages} {1539} (\bibinfo {year} {2010})}\BibitemShut {NoStop}%
\bibitem [{\citenamefont {Chang}\ and\ \citenamefont
  {Niu}(1996)}]{Chang1995PRB}%
  \BibitemOpen
  \bibfield  {author} {\bibinfo {author} {\bibfnamefont {M.-C.}\ \bibnamefont
  {Chang}}\ and\ \bibinfo {author} {\bibfnamefont {Q.}~\bibnamefont {Niu}},\
  }\href {https://doi.org/10.1103/PhysRevB.53.7010} {\bibfield  {journal}
  {\bibinfo  {journal} {Phys. Rev. B}\ }\textbf {\bibinfo {volume} {53}},\
  \bibinfo {pages} {7010} (\bibinfo {year} {1996})}\BibitemShut {NoStop}%
\bibitem [{\citenamefont {Chang}\ and\ \citenamefont
  {Niu}(1995)}]{Chang1995PRL}%
  \BibitemOpen
  \bibfield  {author} {\bibinfo {author} {\bibfnamefont {M.-C.}\ \bibnamefont
  {Chang}}\ and\ \bibinfo {author} {\bibfnamefont {Q.}~\bibnamefont {Niu}},\
  }\href {https://doi.org/10.1103/PhysRevLett.75.1348} {\bibfield  {journal}
  {\bibinfo  {journal} {Phys. Rev. Lett.}\ }\textbf {\bibinfo {volume} {75}},\
  \bibinfo {pages} {1348} (\bibinfo {year} {1995})}\BibitemShut {NoStop}%
\bibitem [{\citenamefont {Sundaram}\ and\ \citenamefont
  {Niu}(1999)}]{Sundaram1999}%
  \BibitemOpen
  \bibfield  {author} {\bibinfo {author} {\bibfnamefont {G.}~\bibnamefont
  {Sundaram}}\ and\ \bibinfo {author} {\bibfnamefont {Q.}~\bibnamefont {Niu}},\
  }\href {https://doi.org/10.1103/PhysRevB.59.14915} {\bibfield  {journal}
  {\bibinfo  {journal} {Phys. Rev. B}\ }\textbf {\bibinfo {volume} {59}},\
  \bibinfo {pages} {14915} (\bibinfo {year} {1999})}\BibitemShut {NoStop}%
\bibitem [{\citenamefont {Manchon}\ \emph {et~al.}(2019)\citenamefont
  {Manchon}, \citenamefont {\ifmmode~\check{Z}\else \v{Z}\fi{}elezn\'y},
  \citenamefont {Miron}, \citenamefont {Jungwirth}, \citenamefont {Sinova},
  \citenamefont {Thiaville}, \citenamefont {Garello},\ and\ \citenamefont
  {Gambardella}}]{Manchon2019}%
  \BibitemOpen
  \bibfield  {author} {\bibinfo {author} {\bibfnamefont {A.}~\bibnamefont
  {Manchon}}, \bibinfo {author} {\bibfnamefont {J.}~\bibnamefont
  {\ifmmode~\check{Z}\else \v{Z}\fi{}elezn\'y}}, \bibinfo {author}
  {\bibfnamefont {I.~M.}\ \bibnamefont {Miron}}, \bibinfo {author}
  {\bibfnamefont {T.}~\bibnamefont {Jungwirth}}, \bibinfo {author}
  {\bibfnamefont {J.}~\bibnamefont {Sinova}}, \bibinfo {author} {\bibfnamefont
  {A.}~\bibnamefont {Thiaville}}, \bibinfo {author} {\bibfnamefont
  {K.}~\bibnamefont {Garello}},\ and\ \bibinfo {author} {\bibfnamefont
  {P.}~\bibnamefont {Gambardella}},\ }\href
  {https://doi.org/10.1103/RevModPhys.91.035004} {\bibfield  {journal}
  {\bibinfo  {journal} {Rev. Mod. Phys.}\ }\textbf {\bibinfo {volume} {91}},\
  \bibinfo {pages} {035004} (\bibinfo {year} {2019})}\BibitemShut {NoStop}%
\bibitem [{\citenamefont {Zhu}\ \emph {et~al.}(2019)\citenamefont {Zhu},
  \citenamefont {Ralph},\ and\ \citenamefont {Buhrman}}]{Zhu2019}%
  \BibitemOpen
  \bibfield  {author} {\bibinfo {author} {\bibfnamefont {L.}~\bibnamefont
  {Zhu}}, \bibinfo {author} {\bibfnamefont {D.~C.}\ \bibnamefont {Ralph}},\
  and\ \bibinfo {author} {\bibfnamefont {R.~A.}\ \bibnamefont {Buhrman}},\
  }\href {https://doi.org/10.1103/PhysRevLett.122.077201} {\bibfield  {journal}
  {\bibinfo  {journal} {Phys. Rev. Lett.}\ }\textbf {\bibinfo {volume} {122}},\
  \bibinfo {pages} {077201} (\bibinfo {year} {2019})}\BibitemShut {NoStop}%
\bibitem [{\citenamefont {Gueckstock}\ \emph {et~al.}(2021)\citenamefont
  {Gueckstock}, \citenamefont {N\'{a}dvorn\'{i}k}, \citenamefont {Gradhand},
  \citenamefont {Seifert}, \citenamefont {Bierhance}, \citenamefont {Rouzegar},
  \citenamefont {Wolf}, \citenamefont {Vafaee}, \citenamefont {Cramer},
  \citenamefont {Syskaki}, \citenamefont {Woltersdorf}, \citenamefont {Mertig},
  \citenamefont {Jakob}, \citenamefont {Kl\"{a}ui},\ and\ \citenamefont
  {Kampfrath}}]{Gueckstock2021}%
  \BibitemOpen
  \bibfield  {author} {\bibinfo {author} {\bibfnamefont {O.}~\bibnamefont
  {Gueckstock}}, \bibinfo {author} {\bibfnamefont {L.}~\bibnamefont
  {N\'{a}dvorn\'{i}k}}, \bibinfo {author} {\bibfnamefont {M.}~\bibnamefont
  {Gradhand}}, \bibinfo {author} {\bibfnamefont {T.~S.}\ \bibnamefont
  {Seifert}}, \bibinfo {author} {\bibfnamefont {G.}~\bibnamefont {Bierhance}},
  \bibinfo {author} {\bibfnamefont {R.}~\bibnamefont {Rouzegar}}, \bibinfo
  {author} {\bibfnamefont {M.}~\bibnamefont {Wolf}}, \bibinfo {author}
  {\bibfnamefont {M.}~\bibnamefont {Vafaee}}, \bibinfo {author} {\bibfnamefont
  {J.}~\bibnamefont {Cramer}}, \bibinfo {author} {\bibfnamefont {M.~A.}\
  \bibnamefont {Syskaki}}, \bibinfo {author} {\bibfnamefont {G.}~\bibnamefont
  {Woltersdorf}}, \bibinfo {author} {\bibfnamefont {I.}~\bibnamefont {Mertig}},
  \bibinfo {author} {\bibfnamefont {G.}~\bibnamefont {Jakob}}, \bibinfo
  {author} {\bibfnamefont {M.}~\bibnamefont {Kl\"{a}ui}},\ and\ \bibinfo
  {author} {\bibfnamefont {T.}~\bibnamefont {Kampfrath}},\ }\href
  {https://doi.org/https://doi.org/10.1002/adma.202006281} {\bibfield
  {journal} {\bibinfo  {journal} {Adv. Mater.}\ }\textbf {\bibinfo {volume}
  {33}},\ \bibinfo {pages} {2006281} (\bibinfo {year} {2021})}\BibitemShut
  {NoStop}%
\bibitem [{\citenamefont {Nikoli\ifmmode~\acute{c}\else \'{c}\fi{}}\ \emph
  {et~al.}(2005)\citenamefont {Nikoli\ifmmode~\acute{c}\else \'{c}\fi{}},
  \citenamefont {Souma}, \citenamefont {Z\^arbo},\ and\ \citenamefont
  {Sinova}}]{Nikolic2005}%
  \BibitemOpen
  \bibfield  {author} {\bibinfo {author} {\bibfnamefont {B.~K.}\ \bibnamefont
  {Nikoli\ifmmode~\acute{c}\else \'{c}\fi{}}}, \bibinfo {author} {\bibfnamefont
  {S.}~\bibnamefont {Souma}}, \bibinfo {author} {\bibfnamefont {L.~P.}\
  \bibnamefont {Z\^arbo}},\ and\ \bibinfo {author} {\bibfnamefont
  {J.}~\bibnamefont {Sinova}},\ }\href
  {https://doi.org/10.1103/PhysRevLett.95.046601} {\bibfield  {journal}
  {\bibinfo  {journal} {Phys. Rev. Lett.}\ }\textbf {\bibinfo {volume} {95}},\
  \bibinfo {pages} {046601} (\bibinfo {year} {2005})}\BibitemShut {NoStop}%
\bibitem [{\citenamefont {Nikoli\ifmmode~\acute{c}\else \'{c}\fi{}}\ \emph
  {et~al.}(2006)\citenamefont {Nikoli\ifmmode~\acute{c}\else \'{c}\fi{}},
  \citenamefont {Z\^arbo},\ and\ \citenamefont {Souma}}]{Nikolic2006}%
  \BibitemOpen
  \bibfield  {author} {\bibinfo {author} {\bibfnamefont {B.~K.}\ \bibnamefont
  {Nikoli\ifmmode~\acute{c}\else \'{c}\fi{}}}, \bibinfo {author} {\bibfnamefont
  {L.~P.}\ \bibnamefont {Z\^arbo}},\ and\ \bibinfo {author} {\bibfnamefont
  {S.}~\bibnamefont {Souma}},\ }\href
  {https://doi.org/10.1103/PhysRevB.73.075303} {\bibfield  {journal} {\bibinfo
  {journal} {Phys. Rev. B}\ }\textbf {\bibinfo {volume} {73}},\ \bibinfo
  {pages} {075303} (\bibinfo {year} {2006})}\BibitemShut {NoStop}%
\bibitem [{\citenamefont {G\'eranton}\ \emph {et~al.}(2016)\citenamefont
  {G\'eranton}, \citenamefont {Zimmermann}, \citenamefont {Long}, \citenamefont
  {Mavropoulos}, \citenamefont {Bl\"ugel}, \citenamefont {Freimuth},\ and\
  \citenamefont {Mokrousov}}]{Geranton2016}%
  \BibitemOpen
  \bibfield  {author} {\bibinfo {author} {\bibfnamefont {G.}~\bibnamefont
  {G\'eranton}}, \bibinfo {author} {\bibfnamefont {B.}~\bibnamefont
  {Zimmermann}}, \bibinfo {author} {\bibfnamefont {N.~H.}\ \bibnamefont
  {Long}}, \bibinfo {author} {\bibfnamefont {P.}~\bibnamefont {Mavropoulos}},
  \bibinfo {author} {\bibfnamefont {S.}~\bibnamefont {Bl\"ugel}}, \bibinfo
  {author} {\bibfnamefont {F.}~\bibnamefont {Freimuth}},\ and\ \bibinfo
  {author} {\bibfnamefont {Y.}~\bibnamefont {Mokrousov}},\ }\href
  {https://doi.org/10.1103/PhysRevB.93.224420} {\bibfield  {journal} {\bibinfo
  {journal} {Phys. Rev. B}\ }\textbf {\bibinfo {volume} {93}},\ \bibinfo
  {pages} {224420} (\bibinfo {year} {2016})}\BibitemShut {NoStop}%
\bibitem [{\citenamefont {Fabian}\ \emph {et~al.}(2021)\citenamefont {Fabian},
  \citenamefont {Czerner}, \citenamefont {Heiliger}, \citenamefont {Rossignol},
  \citenamefont {Wu},\ and\ \citenamefont {Gradhand}}]{Fabian2021}%
  \BibitemOpen
  \bibfield  {author} {\bibinfo {author} {\bibfnamefont {A.}~\bibnamefont
  {Fabian}}, \bibinfo {author} {\bibfnamefont {M.}~\bibnamefont {Czerner}},
  \bibinfo {author} {\bibfnamefont {C.}~\bibnamefont {Heiliger}}, \bibinfo
  {author} {\bibfnamefont {H.}~\bibnamefont {Rossignol}}, \bibinfo {author}
  {\bibfnamefont {M.-H.}\ \bibnamefont {Wu}},\ and\ \bibinfo {author}
  {\bibfnamefont {M.}~\bibnamefont {Gradhand}},\ }\href
  {https://doi.org/10.1103/PhysRevB.104.054402} {\bibfield  {journal} {\bibinfo
   {journal} {Phys. Rev. B}\ }\textbf {\bibinfo {volume} {104}},\ \bibinfo
  {pages} {054402} (\bibinfo {year} {2021})}\BibitemShut {NoStop}%
\bibitem [{\citenamefont {Kimura}\ \emph {et~al.}(2007)\citenamefont {Kimura},
  \citenamefont {Otani}, \citenamefont {Sato}, \citenamefont {Takahashi},\ and\
  \citenamefont {Maekawa}}]{Kimura2007}%
  \BibitemOpen
  \bibfield  {author} {\bibinfo {author} {\bibfnamefont {T.}~\bibnamefont
  {Kimura}}, \bibinfo {author} {\bibfnamefont {Y.}~\bibnamefont {Otani}},
  \bibinfo {author} {\bibfnamefont {T.}~\bibnamefont {Sato}}, \bibinfo {author}
  {\bibfnamefont {S.}~\bibnamefont {Takahashi}},\ and\ \bibinfo {author}
  {\bibfnamefont {S.}~\bibnamefont {Maekawa}},\ }\href
  {https://doi.org/10.1103/PhysRevLett.98.156601} {\bibfield  {journal}
  {\bibinfo  {journal} {Phys. Rev. Lett.}\ }\textbf {\bibinfo {volume} {98}},\
  \bibinfo {pages} {156601} (\bibinfo {year} {2007})}\BibitemShut {NoStop}%
\bibitem [{\citenamefont {Gradhand}\ \emph
  {et~al.}(2010{\natexlab{a}})\citenamefont {Gradhand}, \citenamefont
  {Fedorov}, \citenamefont {Zahn},\ and\ \citenamefont
  {Mertig}}]{Gradhand2010_PRB020403}%
  \BibitemOpen
  \bibfield  {author} {\bibinfo {author} {\bibfnamefont {M.}~\bibnamefont
  {Gradhand}}, \bibinfo {author} {\bibfnamefont {D.~V.}\ \bibnamefont
  {Fedorov}}, \bibinfo {author} {\bibfnamefont {P.}~\bibnamefont {Zahn}},\ and\
  \bibinfo {author} {\bibfnamefont {I.}~\bibnamefont {Mertig}},\ }\href
  {https://doi.org/10.1103/PhysRevB.81.020403} {\bibfield  {journal} {\bibinfo
  {journal} {Phys. Rev. B}\ }\textbf {\bibinfo {volume} {81}},\ \bibinfo
  {pages} {020403(R)} (\bibinfo {year} {2010}{\natexlab{a}})}\BibitemShut
  {NoStop}%
\bibitem [{\citenamefont {Gradhand}\ \emph
  {et~al.}(2010{\natexlab{b}})\citenamefont {Gradhand}, \citenamefont
  {Fedorov}, \citenamefont {Zahn},\ and\ \citenamefont
  {Mertig}}]{Gradhand2010PRL}%
  \BibitemOpen
  \bibfield  {author} {\bibinfo {author} {\bibfnamefont {M.}~\bibnamefont
  {Gradhand}}, \bibinfo {author} {\bibfnamefont {D.~V.}\ \bibnamefont
  {Fedorov}}, \bibinfo {author} {\bibfnamefont {P.}~\bibnamefont {Zahn}},\ and\
  \bibinfo {author} {\bibfnamefont {I.}~\bibnamefont {Mertig}},\ }\href
  {https://doi.org/10.1103/PhysRevLett.104.186403} {\bibfield  {journal}
  {\bibinfo  {journal} {Phys. Rev. Lett.}\ }\textbf {\bibinfo {volume} {104}},\
  \bibinfo {pages} {186403} (\bibinfo {year} {2010}{\natexlab{b}})}\BibitemShut
  {NoStop}%
\bibitem [{\citenamefont {Avci}\ \emph {et~al.}(2014)\citenamefont {Avci},
  \citenamefont {Garello}, \citenamefont {Nistor}, \citenamefont {Godey},
  \citenamefont {Ballesteros}, \citenamefont {Mugarza}, \citenamefont {Barla},
  \citenamefont {Valvidares}, \citenamefont {Pellegrin}, \citenamefont {Ghosh},
  \citenamefont {Miron}, \citenamefont {Boulle}, \citenamefont {Auffret},
  \citenamefont {Gaudin},\ and\ \citenamefont {Gambardella}}]{Avci2014}%
  \BibitemOpen
  \bibfield  {author} {\bibinfo {author} {\bibfnamefont {C.~O.}\ \bibnamefont
  {Avci}}, \bibinfo {author} {\bibfnamefont {K.}~\bibnamefont {Garello}},
  \bibinfo {author} {\bibfnamefont {C.}~\bibnamefont {Nistor}}, \bibinfo
  {author} {\bibfnamefont {S.}~\bibnamefont {Godey}}, \bibinfo {author}
  {\bibfnamefont {B.}~\bibnamefont {Ballesteros}}, \bibinfo {author}
  {\bibfnamefont {A.}~\bibnamefont {Mugarza}}, \bibinfo {author} {\bibfnamefont
  {A.}~\bibnamefont {Barla}}, \bibinfo {author} {\bibfnamefont
  {M.}~\bibnamefont {Valvidares}}, \bibinfo {author} {\bibfnamefont
  {E.}~\bibnamefont {Pellegrin}}, \bibinfo {author} {\bibfnamefont
  {A.}~\bibnamefont {Ghosh}}, \bibinfo {author} {\bibfnamefont {I.~M.}\
  \bibnamefont {Miron}}, \bibinfo {author} {\bibfnamefont {O.}~\bibnamefont
  {Boulle}}, \bibinfo {author} {\bibfnamefont {S.}~\bibnamefont {Auffret}},
  \bibinfo {author} {\bibfnamefont {G.}~\bibnamefont {Gaudin}},\ and\ \bibinfo
  {author} {\bibfnamefont {P.}~\bibnamefont {Gambardella}},\ }\href
  {https://doi.org/10.1103/PhysRevB.89.214419} {\bibfield  {journal} {\bibinfo
  {journal} {Phys. Rev. B}\ }\textbf {\bibinfo {volume} {89}},\ \bibinfo
  {pages} {214419} (\bibinfo {year} {2014})}\BibitemShut {NoStop}%
\bibitem [{\citenamefont {Herschbach}\ \emph {et~al.}(2014)\citenamefont
  {Herschbach}, \citenamefont {Fedorov}, \citenamefont {Gradhand},\ and\
  \citenamefont {Mertig}}]{Herschbach2014}%
  \BibitemOpen
  \bibfield  {author} {\bibinfo {author} {\bibfnamefont {C.}~\bibnamefont
  {Herschbach}}, \bibinfo {author} {\bibfnamefont {D.~V.}\ \bibnamefont
  {Fedorov}}, \bibinfo {author} {\bibfnamefont {M.}~\bibnamefont {Gradhand}},\
  and\ \bibinfo {author} {\bibfnamefont {I.}~\bibnamefont {Mertig}},\ }\href
  {https://doi.org/10.1103/PhysRevB.90.180406} {\bibfield  {journal} {\bibinfo
  {journal} {Phys. Rev. B}\ }\textbf {\bibinfo {volume} {90}},\ \bibinfo
  {pages} {180406(R)} (\bibinfo {year} {2014})}\BibitemShut {NoStop}%
\bibitem [{\citenamefont {Saidaoui}\ and\ \citenamefont
  {Manchon}(2016)}]{Saidaoui2016}%
  \BibitemOpen
  \bibfield  {author} {\bibinfo {author} {\bibfnamefont {Hamed Ben Mohamed}\
  \bibnamefont {Saidaoui}}\ and\ \bibinfo {author} {\bibfnamefont
  {A.}~\bibnamefont {Manchon}},\ }\href
  {https://doi.org/10.1103/PhysRevLett.117.036601} {\bibfield  {journal}
  {\bibinfo  {journal} {Phys. Rev. Lett.}\ }\textbf {\bibinfo {volume} {117}},\
  \bibinfo {pages} {036601} (\bibinfo {year} {2016})}\BibitemShut {NoStop}%
\bibitem [{\citenamefont {Freimuth}\ \emph {et~al.}(2014)\citenamefont
  {Freimuth}, \citenamefont {Bl\"ugel},\ and\ \citenamefont
  {Mokrousov}}]{Freimuth2014}%
  \BibitemOpen
  \bibfield  {author} {\bibinfo {author} {\bibfnamefont {F.}~\bibnamefont
  {Freimuth}}, \bibinfo {author} {\bibfnamefont {S.}~\bibnamefont {Bl\"ugel}},\
  and\ \bibinfo {author} {\bibfnamefont {Y.}~\bibnamefont {Mokrousov}},\ }\href
  {https://doi.org/10.1103/PhysRevB.90.174423} {\bibfield  {journal} {\bibinfo
  {journal} {Phys. Rev. B}\ }\textbf {\bibinfo {volume} {90}},\ \bibinfo
  {pages} {174423} (\bibinfo {year} {2014})}\BibitemShut {NoStop}%
\bibitem [{\citenamefont {Garello}\ \emph {et~al.}(2013)\citenamefont
  {Garello}, \citenamefont {Miron}, \citenamefont {Avci}, \citenamefont
  {Freimuth}, \citenamefont {Mokrousov}, \citenamefont {Bl{\"u}gel},
  \citenamefont {Auffret}, \citenamefont {Boulle}, \citenamefont {Gaudin},\
  and\ \citenamefont {Gambardella}}]{Garello2013}%
  \BibitemOpen
  \bibfield  {author} {\bibinfo {author} {\bibfnamefont {K.}~\bibnamefont
  {Garello}}, \bibinfo {author} {\bibfnamefont {I.~M.}\ \bibnamefont {Miron}},
  \bibinfo {author} {\bibfnamefont {C.~O.}\ \bibnamefont {Avci}}, \bibinfo
  {author} {\bibfnamefont {F.}~\bibnamefont {Freimuth}}, \bibinfo {author}
  {\bibfnamefont {Y.}~\bibnamefont {Mokrousov}}, \bibinfo {author}
  {\bibfnamefont {S.}~\bibnamefont {Bl{\"u}gel}}, \bibinfo {author}
  {\bibfnamefont {S.}~\bibnamefont {Auffret}}, \bibinfo {author} {\bibfnamefont
  {O.}~\bibnamefont {Boulle}}, \bibinfo {author} {\bibfnamefont
  {G.}~\bibnamefont {Gaudin}},\ and\ \bibinfo {author} {\bibfnamefont
  {P.}~\bibnamefont {Gambardella}},\ }\href
  {https://doi.org/10.1038/nnano.2013.145} {\bibfield  {journal} {\bibinfo
  {journal} {Nature Nanotechnology}\ }\textbf {\bibinfo {volume} {8}},\
  \bibinfo {pages} {587} (\bibinfo {year} {2013})}\BibitemShut {NoStop}%
\bibitem [{\citenamefont {Gradhand}\ \emph
  {et~al.}(2010{\natexlab{c}})\citenamefont {Gradhand}, \citenamefont
  {Fedorov}, \citenamefont {Zahn},\ and\ \citenamefont
  {Mertig}}]{Gradhand2010_PRB245109}%
  \BibitemOpen
  \bibfield  {author} {\bibinfo {author} {\bibfnamefont {M.}~\bibnamefont
  {Gradhand}}, \bibinfo {author} {\bibfnamefont {D.~V.}\ \bibnamefont
  {Fedorov}}, \bibinfo {author} {\bibfnamefont {P.}~\bibnamefont {Zahn}},\ and\
  \bibinfo {author} {\bibfnamefont {I.}~\bibnamefont {Mertig}},\ }\href
  {https://doi.org/10.1103/PhysRevB.81.245109} {\bibfield  {journal} {\bibinfo
  {journal} {Phys. Rev. B}\ }\textbf {\bibinfo {volume} {81}},\ \bibinfo
  {pages} {245109} (\bibinfo {year} {2010}{\natexlab{c}})}\BibitemShut
  {NoStop}%
\bibitem [{\citenamefont {Korringa}(1947)}]{KORRINGA1947}%
  \BibitemOpen
  \bibfield  {author} {\bibinfo {author} {\bibfnamefont {J.}~\bibnamefont
  {Korringa}},\ }\href
  {https://doi.org/https://doi.org/10.1016/0031-8914(47)90013-X} {\bibfield
  {journal} {\bibinfo  {journal} {Physica}\ }\textbf {\bibinfo {volume} {13}},\
  \bibinfo {pages} {392 } (\bibinfo {year} {1947})}\BibitemShut {NoStop}%
\bibitem [{\citenamefont {Kohn}\ and\ \citenamefont
  {Rostoker}(1954)}]{Kohn1954}%
  \BibitemOpen
  \bibfield  {author} {\bibinfo {author} {\bibfnamefont {W.}~\bibnamefont
  {Kohn}}\ and\ \bibinfo {author} {\bibfnamefont {N.}~\bibnamefont
  {Rostoker}},\ }\href {https://doi.org/10.1103/PhysRev.94.1111} {\bibfield
  {journal} {\bibinfo  {journal} {Phys. Rev.}\ }\textbf {\bibinfo {volume}
  {94}},\ \bibinfo {pages} {1111} (\bibinfo {year} {1954})}\BibitemShut
  {NoStop}%
\bibitem [{\citenamefont {Gradhand}\ \emph {et~al.}(2009)\citenamefont
  {Gradhand}, \citenamefont {Czerner}, \citenamefont {Fedorov}, \citenamefont
  {Zahn}, \citenamefont {Yavorsky}, \citenamefont {Szunyogh},\ and\
  \citenamefont {Mertig}}]{Gradhand2009}%
  \BibitemOpen
  \bibfield  {author} {\bibinfo {author} {\bibfnamefont {M.}~\bibnamefont
  {Gradhand}}, \bibinfo {author} {\bibfnamefont {M.}~\bibnamefont {Czerner}},
  \bibinfo {author} {\bibfnamefont {D.~V.}\ \bibnamefont {Fedorov}}, \bibinfo
  {author} {\bibfnamefont {P.}~\bibnamefont {Zahn}}, \bibinfo {author}
  {\bibfnamefont {B.~Y.}\ \bibnamefont {Yavorsky}}, \bibinfo {author}
  {\bibfnamefont {L.}~\bibnamefont {Szunyogh}},\ and\ \bibinfo {author}
  {\bibfnamefont {I.}~\bibnamefont {Mertig}},\ }\href
  {https://doi.org/10.1103/PhysRevB.80.224413} {\bibfield  {journal} {\bibinfo
  {journal} {Phys. Rev. B}\ }\textbf {\bibinfo {volume} {80}},\ \bibinfo
  {pages} {224413} (\bibinfo {year} {2009})}\BibitemShut {NoStop}%
\bibitem [{\citenamefont {Sagasta}\ \emph {et~al.}(2016)\citenamefont
  {Sagasta}, \citenamefont {Omori}, \citenamefont {Isasa}, \citenamefont
  {Gradhand}, \citenamefont {Hueso}, \citenamefont {Niimi}, \citenamefont
  {Otani},\ and\ \citenamefont {Casanova}}]{Sagasta2016}%
  \BibitemOpen
  \bibfield  {author} {\bibinfo {author} {\bibfnamefont {E.}~\bibnamefont
  {Sagasta}}, \bibinfo {author} {\bibfnamefont {Y.}~\bibnamefont {Omori}},
  \bibinfo {author} {\bibfnamefont {M.}~\bibnamefont {Isasa}}, \bibinfo
  {author} {\bibfnamefont {M.}~\bibnamefont {Gradhand}}, \bibinfo {author}
  {\bibfnamefont {L.~E.}\ \bibnamefont {Hueso}}, \bibinfo {author}
  {\bibfnamefont {Y.}~\bibnamefont {Niimi}}, \bibinfo {author} {\bibfnamefont
  {Y.C.}~\bibnamefont {Otani}},\ and\ \bibinfo {author} {\bibfnamefont
  {F.}~\bibnamefont {Casanova}},\ }\href
  {https://doi.org/10.1103/PhysRevB.94.060412} {\bibfield  {journal} {\bibinfo
  {journal} {Phys. Rev. B}\ }\textbf {\bibinfo {volume} {94}},\ \bibinfo
  {pages} {060412(R)} (\bibinfo {year} {2016})}\BibitemShut {NoStop}%
\bibitem [{\citenamefont {Liu}\ \emph {et~al.}(2011)\citenamefont {Liu},
  \citenamefont {Moriyama}, \citenamefont {Ralph},\ and\ \citenamefont
  {Buhrman}}]{Liu2011_PRL}%
  \BibitemOpen
  \bibfield  {author} {\bibinfo {author} {\bibfnamefont {L.}~\bibnamefont
  {Liu}}, \bibinfo {author} {\bibfnamefont {T.}~\bibnamefont {Moriyama}},
  \bibinfo {author} {\bibfnamefont {D.~C.}\ \bibnamefont {Ralph}},\ and\
  \bibinfo {author} {\bibfnamefont {R.~A.}\ \bibnamefont {Buhrman}},\ }\href
  {https://doi.org/10.1103/PhysRevLett.106.036601} {\bibfield  {journal}
  {\bibinfo  {journal} {Phys. Rev. Lett.}\ }\textbf {\bibinfo {volume} {106}},\
  \bibinfo {pages} {036601} (\bibinfo {year} {2011})}\BibitemShut {NoStop}%
\bibitem [{\citenamefont {Ando}\ \emph {et~al.}(2008)\citenamefont {Ando},
  \citenamefont {Takahashi}, \citenamefont {Harii}, \citenamefont {Sasage},
  \citenamefont {Ieda}, \citenamefont {Maekawa},\ and\ \citenamefont
  {Saitoh}}]{Ando2011_PRL}%
  \BibitemOpen
  \bibfield  {author} {\bibinfo {author} {\bibfnamefont {K.}~\bibnamefont
  {Ando}}, \bibinfo {author} {\bibfnamefont {S.}~\bibnamefont {Takahashi}},
  \bibinfo {author} {\bibfnamefont {K.}~\bibnamefont {Harii}}, \bibinfo
  {author} {\bibfnamefont {K.}~\bibnamefont {Sasage}}, \bibinfo {author}
  {\bibfnamefont {J.}~\bibnamefont {Ieda}}, \bibinfo {author} {\bibfnamefont
  {S.}~\bibnamefont {Maekawa}},\ and\ \bibinfo {author} {\bibfnamefont
  {E.}~\bibnamefont {Saitoh}},\ }\href
  {https://doi.org/10.1103/PhysRevLett.101.036601} {\bibfield  {journal}
  {\bibinfo  {journal} {Phys. Rev. Lett.}\ }\textbf {\bibinfo {volume} {101}},\
  \bibinfo {pages} {036601} (\bibinfo {year} {2008})}\BibitemShut {NoStop}%
\bibitem [{\citenamefont {Sagasta}\ \emph {et~al.}(2017)\citenamefont
  {Sagasta}, \citenamefont {Omori}, \citenamefont {Isasa}, \citenamefont
  {Otani}, \citenamefont {Hueso},\ and\ \citenamefont
  {Casanova}}]{Sagasta2017}%
  \BibitemOpen
  \bibfield  {author} {\bibinfo {author} {\bibfnamefont {E.}~\bibnamefont
  {Sagasta}}, \bibinfo {author} {\bibfnamefont {Y.}~\bibnamefont {Omori}},
  \bibinfo {author} {\bibfnamefont {M.}~\bibnamefont {Isasa}}, \bibinfo
  {author} {\bibfnamefont {Y.}~\bibnamefont {Otani}}, \bibinfo {author}
  {\bibfnamefont {L.~E.}\ \bibnamefont {Hueso}},\ and\ \bibinfo {author}
  {\bibfnamefont {F.}~\bibnamefont {Casanova}},\ }\href
  {https://doi.org/10.1063/1.4990652} {\bibfield  {journal} {\bibinfo
  {journal} {Appl. Phys. Lett.}\ }\textbf {\bibinfo {volume} {111}},\ \bibinfo
  {pages} {082407} (\bibinfo {year} {2017})}\BibitemShut {NoStop}%
\bibitem [{\citenamefont {Kimura}\ \emph {et~al.}(2005)\citenamefont {Kimura},
  \citenamefont {Hamrle},\ and\ \citenamefont {Otani}}]{Kimura2005}%
  \BibitemOpen
  \bibfield  {author} {\bibinfo {author} {\bibfnamefont {T.}~\bibnamefont
  {Kimura}}, \bibinfo {author} {\bibfnamefont {J.}~\bibnamefont {Hamrle}},\
  and\ \bibinfo {author} {\bibfnamefont {Y.}~\bibnamefont {Otani}},\ }\href
  {https://doi.org/10.1103/PhysRevB.72.014461} {\bibfield  {journal} {\bibinfo
  {journal} {Phys. Rev. B}\ }\textbf {\bibinfo {volume} {72}},\ \bibinfo
  {pages} {014461} (\bibinfo {year} {2005})}\BibitemShut {NoStop}%
\bibitem [{\citenamefont {Zhang}\ \emph {et~al.}(2013)\citenamefont {Zhang},
  \citenamefont {Vlaminck}, \citenamefont {Pearson}, \citenamefont {Divan},
  \citenamefont {Bader},\ and\ \citenamefont {Hoffmann}}]{Zhang2013}%
  \BibitemOpen
  \bibfield  {author} {\bibinfo {author} {\bibfnamefont {W.}~\bibnamefont
  {Zhang}}, \bibinfo {author} {\bibfnamefont {V.}~\bibnamefont {Vlaminck}},
  \bibinfo {author} {\bibfnamefont {J.~E.}\ \bibnamefont {Pearson}}, \bibinfo
  {author} {\bibfnamefont {R.}~\bibnamefont {Divan}}, \bibinfo {author}
  {\bibfnamefont {S.~D.}\ \bibnamefont {Bader}},\ and\ \bibinfo {author}
  {\bibfnamefont {A.}~\bibnamefont {Hoffmann}},\ }\href
  {https://doi.org/10.1063/1.4848102} {\bibfield  {journal} {\bibinfo
  {journal} {Appl. Phys. Lett.}\ }\textbf {\bibinfo {volume} {103}},\ \bibinfo
  {pages} {242414} (\bibinfo {year} {2013})}\BibitemShut {NoStop}%
\bibitem [{\citenamefont {Guo}\ \emph {et~al.}(2008)\citenamefont {Guo},
  \citenamefont {Murakami}, \citenamefont {Chen},\ and\ \citenamefont
  {Nagaosa}}]{Guo2008}%
  \BibitemOpen
  \bibfield  {author} {\bibinfo {author} {\bibfnamefont {G.~Y.}\ \bibnamefont
  {Guo}}, \bibinfo {author} {\bibfnamefont {S.}~\bibnamefont {Murakami}},
  \bibinfo {author} {\bibfnamefont {T.-W.}\ \bibnamefont {Chen}},\ and\
  \bibinfo {author} {\bibfnamefont {N.}~\bibnamefont {Nagaosa}},\ }\href
  {https://doi.org/10.1103/PhysRevLett.100.096401} {\bibfield  {journal}
  {\bibinfo  {journal} {Phys. Rev. Lett.}\ }\textbf {\bibinfo {volume} {100}},\
  \bibinfo {pages} {096401} (\bibinfo {year} {2008})}\BibitemShut {NoStop}%
\bibitem [{\citenamefont {Gradhand}\ \emph {et~al.}(2011)\citenamefont
  {Gradhand}, \citenamefont {Fedorov}, \citenamefont {Pientka}, \citenamefont
  {Zahn}, \citenamefont {Mertig},\ and\ \citenamefont
  {Gy\"orffy}}]{Gradhand2011}%
  \BibitemOpen
  \bibfield  {author} {\bibinfo {author} {\bibfnamefont {M.}~\bibnamefont
  {Gradhand}}, \bibinfo {author} {\bibfnamefont {D.~V.}\ \bibnamefont
  {Fedorov}}, \bibinfo {author} {\bibfnamefont {F.}~\bibnamefont {Pientka}},
  \bibinfo {author} {\bibfnamefont {P.}~\bibnamefont {Zahn}}, \bibinfo {author}
  {\bibfnamefont {I.}~\bibnamefont {Mertig}},\ and\ \bibinfo {author}
  {\bibfnamefont {B.~L.}\ \bibnamefont {Gy\"orffy}},\ }\href
  {https://doi.org/10.1103/PhysRevB.84.075113} {\bibfield  {journal} {\bibinfo
  {journal} {Phys. Rev. B}\ }\textbf {\bibinfo {volume} {84}},\ \bibinfo
  {pages} {075113} (\bibinfo {year} {2011})}\BibitemShut {NoStop}%
\bibitem [{\citenamefont {Fert}\ and\ \citenamefont {Levy}(2011)}]{Fert2011}%
  \BibitemOpen
  \bibfield  {author} {\bibinfo {author} {\bibfnamefont {A.}~\bibnamefont
  {Fert}}\ and\ \bibinfo {author} {\bibfnamefont {P.~M.}\ \bibnamefont
  {Levy}},\ }\href {https://doi.org/10.1103/PhysRevLett.106.157208} {\bibfield
  {journal} {\bibinfo  {journal} {Phys. Rev. Lett.}\ }\textbf {\bibinfo
  {volume} {106}},\ \bibinfo {pages} {157208} (\bibinfo {year}
  {2011})}\BibitemShut {NoStop}%
\bibitem [{\citenamefont {Seemann}\ \emph {et~al.}(2015)\citenamefont
  {Seemann}, \citenamefont {K\"odderitzsch}, \citenamefont {Wimmer},\ and\
  \citenamefont {Ebert}}]{Seemann2015}%
  \BibitemOpen
  \bibfield  {author} {\bibinfo {author} {\bibfnamefont {M.}~\bibnamefont
  {Seemann}}, \bibinfo {author} {\bibfnamefont {D.}~\bibnamefont
  {K\"odderitzsch}}, \bibinfo {author} {\bibfnamefont {S.}~\bibnamefont
  {Wimmer}},\ and\ \bibinfo {author} {\bibfnamefont {H.}~\bibnamefont
  {Ebert}},\ }\href {https://doi.org/10.1103/PhysRevB.92.155138} {\bibfield
  {journal} {\bibinfo  {journal} {Phys. Rev. B}\ }\textbf {\bibinfo {volume}
  {92}},\ \bibinfo {pages} {155138} (\bibinfo {year} {2015})}\BibitemShut
  {NoStop}%
\bibitem [{\citenamefont {Lowitzer}\ \emph {et~al.}(2011)\citenamefont
  {Lowitzer}, \citenamefont {Gradhand}, \citenamefont {K\"odderitzsch},
  \citenamefont {Fedorov}, \citenamefont {Mertig},\ and\ \citenamefont
  {Ebert}}]{Lowitzer2011}%
  \BibitemOpen
  \bibfield  {author} {\bibinfo {author} {\bibfnamefont {S.}~\bibnamefont
  {Lowitzer}}, \bibinfo {author} {\bibfnamefont {M.}~\bibnamefont {Gradhand}},
  \bibinfo {author} {\bibfnamefont {D.}~\bibnamefont {K\"odderitzsch}},
  \bibinfo {author} {\bibfnamefont {D.~V.}\ \bibnamefont {Fedorov}}, \bibinfo
  {author} {\bibfnamefont {I.}~\bibnamefont {Mertig}},\ and\ \bibinfo {author}
  {\bibfnamefont {H.}~\bibnamefont {Ebert}},\ }\href
  {https://doi.org/10.1103/PhysRevLett.106.056601} {\bibfield  {journal}
  {\bibinfo  {journal} {Phys. Rev. Lett.}\ }\textbf {\bibinfo {volume} {106}},\
  \bibinfo {pages} {056601} (\bibinfo {year} {2011})}\BibitemShut {NoStop}%
\bibitem [{\citenamefont {Zimmermann}\ \emph {et~al.}(2014)\citenamefont
  {Zimmermann}, \citenamefont {Chadova}, \citenamefont {K\"odderitzsch},
  \citenamefont {Bl\"ugel}, \citenamefont {Ebert}, \citenamefont {Fedorov},
  \citenamefont {Long}, \citenamefont {Mavropoulos}, \citenamefont {Mertig},
  \citenamefont {Mokrousov},\ and\ \citenamefont {Gradhand}}]{Zimmermann2014}%
  \BibitemOpen
  \bibfield  {author} {\bibinfo {author} {\bibfnamefont {B.}~\bibnamefont
  {Zimmermann}}, \bibinfo {author} {\bibfnamefont {K.}~\bibnamefont {Chadova}},
  \bibinfo {author} {\bibfnamefont {D.}~\bibnamefont {K\"odderitzsch}},
  \bibinfo {author} {\bibfnamefont {S.}~\bibnamefont {Bl\"ugel}}, \bibinfo
  {author} {\bibfnamefont {H.}~\bibnamefont {Ebert}}, \bibinfo {author}
  {\bibfnamefont {D.~V.}\ \bibnamefont {Fedorov}}, \bibinfo {author}
  {\bibfnamefont {N.~H.}\ \bibnamefont {Long}}, \bibinfo {author}
  {\bibfnamefont {P.}~\bibnamefont {Mavropoulos}}, \bibinfo {author}
  {\bibfnamefont {I.}~\bibnamefont {Mertig}}, \bibinfo {author} {\bibfnamefont
  {Y.}~\bibnamefont {Mokrousov}},\ and\ \bibinfo {author} {\bibfnamefont
  {M.}~\bibnamefont {Gradhand}},\ }\href
  {https://doi.org/10.1103/PhysRevB.90.220403} {\bibfield  {journal} {\bibinfo
  {journal} {Phys. Rev. B}\ }\textbf {\bibinfo {volume} {90}},\ \bibinfo
  {pages} {220403(R)} (\bibinfo {year} {2014})}\BibitemShut {NoStop}%
\bibitem [{\citenamefont {Lowitzer}\ \emph {et~al.}(2010)\citenamefont
  {Lowitzer}, \citenamefont {K\"odderitzsch},\ and\ \citenamefont
  {Ebert}}]{Lowitzer2010}%
  \BibitemOpen
  \bibfield  {author} {\bibinfo {author} {\bibfnamefont {S.}~\bibnamefont
  {Lowitzer}}, \bibinfo {author} {\bibfnamefont {D.}~\bibnamefont
  {K\"odderitzsch}},\ and\ \bibinfo {author} {\bibfnamefont {H.}~\bibnamefont
  {Ebert}},\ }\href {https://doi.org/10.1103/PhysRevLett.105.266604} {\bibfield
   {journal} {\bibinfo  {journal} {Phys. Rev. Lett.}\ }\textbf {\bibinfo
  {volume} {105}},\ \bibinfo {pages} {266604} (\bibinfo {year}
  {2010})}\BibitemShut {NoStop}%
\bibitem [{\citenamefont {Freimuth}\ \emph {et~al.}(2010)\citenamefont
  {Freimuth}, \citenamefont {Bl\"ugel},\ and\ \citenamefont
  {Mokrousov}}]{Freimuth2010}%
  \BibitemOpen
  \bibfield  {author} {\bibinfo {author} {\bibfnamefont {F.}~\bibnamefont
  {Freimuth}}, \bibinfo {author} {\bibfnamefont {S.}~\bibnamefont {Bl\"ugel}},\
  and\ \bibinfo {author} {\bibfnamefont {Y.}~\bibnamefont {Mokrousov}},\ }\href
  {https://doi.org/10.1103/PhysRevLett.105.246602} {\bibfield  {journal}
  {\bibinfo  {journal} {Phys. Rev. Lett.}\ }\textbf {\bibinfo {volume} {105}},\
  \bibinfo {pages} {246602} (\bibinfo {year} {2010})}\BibitemShut {NoStop}%
\bibitem [{\citenamefont {Stamm}\ \emph {et~al.}(2017)\citenamefont {Stamm},
  \citenamefont {Murer}, \citenamefont {Berritta}, \citenamefont {Feng},
  \citenamefont {Gabureac}, \citenamefont {Oppeneer},\ and\ \citenamefont
  {Gambardella}}]{Stamm2017}%
  \BibitemOpen
  \bibfield  {author} {\bibinfo {author} {\bibfnamefont {C.}~\bibnamefont
  {Stamm}}, \bibinfo {author} {\bibfnamefont {C.}~\bibnamefont {Murer}},
  \bibinfo {author} {\bibfnamefont {M.}~\bibnamefont {Berritta}}, \bibinfo
  {author} {\bibfnamefont {J.}~\bibnamefont {Feng}}, \bibinfo {author}
  {\bibfnamefont {M.}~\bibnamefont {Gabureac}}, \bibinfo {author}
  {\bibfnamefont {P.~M.}\ \bibnamefont {Oppeneer}},\ and\ \bibinfo {author}
  {\bibfnamefont {P.}~\bibnamefont {Gambardella}},\ }\href
  {https://doi.org/10.1103/PhysRevLett.119.087203} {\bibfield  {journal}
  {\bibinfo  {journal} {Phys. Rev. Lett.}\ }\textbf {\bibinfo {volume} {119}},\
  \bibinfo {pages} {087203} (\bibinfo {year} {2017})}\BibitemShut {NoStop}%
\bibitem [{\citenamefont {Stamm}\ \emph {et~al.}(2019)\citenamefont {Stamm},
  \citenamefont {Murer}, \citenamefont {Acremann}, \citenamefont {Baumgartner},
  \citenamefont {Gort}, \citenamefont {D\"aster}, \citenamefont {Kleibert},
  \citenamefont {Garello}, \citenamefont {Feng}, \citenamefont {Gabureac},
  \citenamefont {Chen}, \citenamefont {St\"ohr},\ and\ \citenamefont
  {Gambardella}}]{Stamm2019}%
  \BibitemOpen
  \bibfield  {author} {\bibinfo {author} {\bibfnamefont {C.}~\bibnamefont
  {Stamm}}, \bibinfo {author} {\bibfnamefont {C.}~\bibnamefont {Murer}},
  \bibinfo {author} {\bibfnamefont {Y.}~\bibnamefont {Acremann}}, \bibinfo
  {author} {\bibfnamefont {M.}~\bibnamefont {Baumgartner}}, \bibinfo {author}
  {\bibfnamefont {R.}~\bibnamefont {Gort}}, \bibinfo {author} {\bibfnamefont
  {S.}~\bibnamefont {D\"aster}}, \bibinfo {author} {\bibfnamefont
  {A.}~\bibnamefont {Kleibert}}, \bibinfo {author} {\bibfnamefont
  {K.}~\bibnamefont {Garello}}, \bibinfo {author} {\bibfnamefont
  {J.}~\bibnamefont {Feng}}, \bibinfo {author} {\bibfnamefont {M.}~\bibnamefont
  {Gabureac}}, \bibinfo {author} {\bibfnamefont {Z.}~\bibnamefont {Chen}},
  \bibinfo {author} {\bibfnamefont {J.}~\bibnamefont {St\"ohr}},\ and\ \bibinfo
  {author} {\bibfnamefont {P.}~\bibnamefont {Gambardella}},\ }\href
  {https://doi.org/10.1103/PhysRevB.100.024426} {\bibfield  {journal} {\bibinfo
   {journal} {Phys. Rev. B}\ }\textbf {\bibinfo {volume} {100}},\ \bibinfo
  {pages} {024426} (\bibinfo {year} {2019})}\BibitemShut {NoStop}%
\bibitem [{\citenamefont {Ding}\ \emph {et~al.}(2020)\citenamefont {Ding},
  \citenamefont {Zhang}, \citenamefont {Jungfleisch}, \citenamefont {Pearson},
  \citenamefont {Ohldag}, \citenamefont {Novosad},\ and\ \citenamefont
  {Hoffmann}}]{Ding2020}%
  \BibitemOpen
  \bibfield  {author} {\bibinfo {author} {\bibfnamefont {J.}~\bibnamefont
  {Ding}}, \bibinfo {author} {\bibfnamefont {W.}~\bibnamefont {Zhang}},
  \bibinfo {author} {\bibfnamefont {M.~B.}\ \bibnamefont {Jungfleisch}},
  \bibinfo {author} {\bibfnamefont {J.~E.}\ \bibnamefont {Pearson}}, \bibinfo
  {author} {\bibfnamefont {H.}~\bibnamefont {Ohldag}}, \bibinfo {author}
  {\bibfnamefont {V.}~\bibnamefont {Novosad}},\ and\ \bibinfo {author}
  {\bibfnamefont {A.}~\bibnamefont {Hoffmann}},\ }\href
  {https://doi.org/10.1103/PhysRevResearch.2.013262} {\bibfield  {journal}
  {\bibinfo  {journal} {Phys. Rev. Research}\ }\textbf {\bibinfo {volume}
  {2}},\ \bibinfo {pages} {013262} (\bibinfo {year} {2020})}\BibitemShut
  {NoStop}%
\bibitem [{\citenamefont {Kukreja}\ \emph {et~al.}(2015)\citenamefont
  {Kukreja}, \citenamefont {Bonetti}, \citenamefont {Chen}, \citenamefont
  {Backes}, \citenamefont {Acremann}, \citenamefont {Katine}, \citenamefont
  {Kent}, \citenamefont {D\"urr}, \citenamefont {Ohldag},\ and\ \citenamefont
  {St\"ohr}}]{Kukreja2015}%
  \BibitemOpen
  \bibfield  {author} {\bibinfo {author} {\bibfnamefont {R.}~\bibnamefont
  {Kukreja}}, \bibinfo {author} {\bibfnamefont {S.}~\bibnamefont {Bonetti}},
  \bibinfo {author} {\bibfnamefont {Z.}~\bibnamefont {Chen}}, \bibinfo {author}
  {\bibfnamefont {D.}~\bibnamefont {Backes}}, \bibinfo {author} {\bibfnamefont
  {Y.}~\bibnamefont {Acremann}}, \bibinfo {author} {\bibfnamefont {J.~A.}\
  \bibnamefont {Katine}}, \bibinfo {author} {\bibfnamefont {A.~D.}\
  \bibnamefont {Kent}}, \bibinfo {author} {\bibfnamefont {H.~A.}\ \bibnamefont
  {D\"urr}}, \bibinfo {author} {\bibfnamefont {H.}~\bibnamefont {Ohldag}},\
  and\ \bibinfo {author} {\bibfnamefont {J.}~\bibnamefont {St\"ohr}},\ }\href
  {https://doi.org/10.1103/PhysRevLett.115.096601} {\bibfield  {journal}
  {\bibinfo  {journal} {Phys. Rev. Lett.}\ }\textbf {\bibinfo {volume} {115}},\
  \bibinfo {pages} {096601} (\bibinfo {year} {2015})}\BibitemShut {NoStop}%
\bibitem [{\citenamefont {Niimi}\ \emph {et~al.}(2012)\citenamefont {Niimi},
  \citenamefont {Kawanishi}, \citenamefont {Wei}, \citenamefont {Deranlot},
  \citenamefont {Yang}, \citenamefont {Chshiev}, \citenamefont {Valet},
  \citenamefont {Fert},\ and\ \citenamefont {Otani}}]{Niimi2012}%
  \BibitemOpen
  \bibfield  {author} {\bibinfo {author} {\bibfnamefont {Y.}~\bibnamefont
  {Niimi}}, \bibinfo {author} {\bibfnamefont {Y.}~\bibnamefont {Kawanishi}},
  \bibinfo {author} {\bibfnamefont {D.~H.}\ \bibnamefont {Wei}}, \bibinfo
  {author} {\bibfnamefont {C.}~\bibnamefont {Deranlot}}, \bibinfo {author}
  {\bibfnamefont {H.~X.}\ \bibnamefont {Yang}}, \bibinfo {author}
  {\bibfnamefont {M.}~\bibnamefont {Chshiev}}, \bibinfo {author} {\bibfnamefont
  {T.}~\bibnamefont {Valet}}, \bibinfo {author} {\bibfnamefont
  {A.}~\bibnamefont {Fert}},\ and\ \bibinfo {author} {\bibfnamefont
  {Y.}~\bibnamefont {Otani}},\ }\href
  {https://doi.org/10.1103/PhysRevLett.109.156602} {\bibfield  {journal}
  {\bibinfo  {journal} {Phys. Rev. Lett.}\ }\textbf {\bibinfo {volume} {109}},\
  \bibinfo {pages} {156602} (\bibinfo {year} {2012})}\BibitemShut {NoStop}%
\bibitem [{\citenamefont {Yoo}\ \emph {et~al.}(1998)\citenamefont {Yoo},
  \citenamefont {Cynn},\ and\ \citenamefont {S\"oderlind}}]{Yoo1998}%
  \BibitemOpen
  \bibfield  {author} {\bibinfo {author} {\bibfnamefont {C.-S.}\ \bibnamefont
  {Yoo}}, \bibinfo {author} {\bibfnamefont {H.}~\bibnamefont {Cynn}},\ and\
  \bibinfo {author} {\bibfnamefont {P.}~\bibnamefont {S\"oderlind}},\ }\href
  {https://doi.org/10.1103/PhysRevB.57.10359} {\bibfield  {journal} {\bibinfo
  {journal} {Phys. Rev. B}\ }\textbf {\bibinfo {volume} {57}},\ \bibinfo
  {pages} {10359} (\bibinfo {year} {1998})}\BibitemShut {NoStop}%
\bibitem [{\citenamefont {Barrett}\ \emph {et~al.}(1963)\citenamefont
  {Barrett}, \citenamefont {Mueller},\ and\ \citenamefont
  {Hitterman}}]{Barrett1963}%
  \BibitemOpen
  \bibfield  {author} {\bibinfo {author} {\bibfnamefont {C.~S.}\ \bibnamefont
  {Barrett}}, \bibinfo {author} {\bibfnamefont {M.~H.}\ \bibnamefont
  {Mueller}},\ and\ \bibinfo {author} {\bibfnamefont {R.~L.}\ \bibnamefont
  {Hitterman}},\ }\href {https://doi.org/10.1103/PhysRev.129.625} {\bibfield
  {journal} {\bibinfo  {journal} {Phys. Rev.}\ }\textbf {\bibinfo {volume}
  {129}},\ \bibinfo {pages} {625} (\bibinfo {year} {1963})}\BibitemShut
  {NoStop}%
\bibitem [{\citenamefont {Springell}\ \emph {et~al.}(2008)\citenamefont
  {Springell}, \citenamefont {Detlefs}, \citenamefont {Lander}, \citenamefont
  {Ward}, \citenamefont {Cowley}, \citenamefont {Ling}, \citenamefont {Goetze},
  \citenamefont {Ahuja}, \citenamefont {Luo},\ and\ \citenamefont
  {Johansson}}]{Springell2008}%
  \BibitemOpen
  \bibfield  {author} {\bibinfo {author} {\bibfnamefont {R.}~\bibnamefont
  {Springell}}, \bibinfo {author} {\bibfnamefont {B.}~\bibnamefont {Detlefs}},
  \bibinfo {author} {\bibfnamefont {G.~H.}\ \bibnamefont {Lander}}, \bibinfo
  {author} {\bibfnamefont {R.~C.~C.}\ \bibnamefont {Ward}}, \bibinfo {author}
  {\bibfnamefont {R.~A.}\ \bibnamefont {Cowley}}, \bibinfo {author}
  {\bibfnamefont {N.}~\bibnamefont {Ling}}, \bibinfo {author} {\bibfnamefont
  {W.}~\bibnamefont {Goetze}}, \bibinfo {author} {\bibfnamefont
  {R.}~\bibnamefont {Ahuja}}, \bibinfo {author} {\bibfnamefont
  {W.}~\bibnamefont {Luo}},\ and\ \bibinfo {author} {\bibfnamefont
  {B.}~\bibnamefont {Johansson}},\ }\href
  {https://doi.org/10.1103/PhysRevB.78.193403} {\bibfield  {journal} {\bibinfo
  {journal} {Phys. Rev. B}\ }\textbf {\bibinfo {volume} {78}},\ \bibinfo
  {pages} {193403} (\bibinfo {year} {2008})}\BibitemShut {NoStop}%
\bibitem [{\citenamefont {Wilson}\ and\ \citenamefont
  {Rundle}(1949)}]{Wilson1949}%
  \BibitemOpen
  \bibfield  {author} {\bibinfo {author} {\bibfnamefont {A.~S.}\ \bibnamefont
  {Wilson}}\ and\ \bibinfo {author} {\bibfnamefont {R.~E.}\ \bibnamefont
  {Rundle}},\ }\href@noop {} {\bibfield  {journal} {\bibinfo  {journal} {Acta
  Cryst.}\ }\textbf {\bibinfo {volume} {2}},\ \bibinfo {pages} {126} (\bibinfo
  {year} {1949})}\BibitemShut {NoStop}%
\bibitem [{\citenamefont {Fedorov}\ \emph {et~al.}(2013)\citenamefont
  {Fedorov}, \citenamefont {Herschbach}, \citenamefont {Johansson},
  \citenamefont {Ostanin}, \citenamefont {Mertig}, \citenamefont {Gradhand},
  \citenamefont {Chadova}, \citenamefont {K\"odderitzsch},\ and\ \citenamefont
  {Ebert}}]{Fedorov2013}%
  \BibitemOpen
  \bibfield  {author} {\bibinfo {author} {\bibfnamefont {D.~V.}\ \bibnamefont
  {Fedorov}}, \bibinfo {author} {\bibfnamefont {C.}~\bibnamefont {Herschbach}},
  \bibinfo {author} {\bibfnamefont {A.}~\bibnamefont {Johansson}}, \bibinfo
  {author} {\bibfnamefont {S.}~\bibnamefont {Ostanin}}, \bibinfo {author}
  {\bibfnamefont {I.}~\bibnamefont {Mertig}}, \bibinfo {author} {\bibfnamefont
  {M.}~\bibnamefont {Gradhand}}, \bibinfo {author} {\bibfnamefont
  {K.}~\bibnamefont {Chadova}}, \bibinfo {author} {\bibfnamefont
  {D.}~\bibnamefont {K\"odderitzsch}},\ and\ \bibinfo {author} {\bibfnamefont
  {H.}~\bibnamefont {Ebert}},\ }\href
  {https://doi.org/10.1103/PhysRevB.88.085116} {\bibfield  {journal} {\bibinfo
  {journal} {Phys. Rev. B}\ }\textbf {\bibinfo {volume} {88}},\ \bibinfo
  {pages} {085116} (\bibinfo {year} {2013})}\BibitemShut {NoStop}%
\bibitem [{\citenamefont {Wu}\ \emph {et~al.}(2020)\citenamefont {Wu},
  \citenamefont {Rossignol},\ and\ \citenamefont {Gradhand}}]{Wu2020}%
  \BibitemOpen
  \bibfield  {author} {\bibinfo {author} {\bibfnamefont {M.-H.}\ \bibnamefont
  {Wu}}, \bibinfo {author} {\bibfnamefont {H.}~\bibnamefont {Rossignol}},\ and\
  \bibinfo {author} {\bibfnamefont {M.}~\bibnamefont {Gradhand}},\ }\href
  {https://doi.org/10.1103/PhysRevB.101.224411} {\bibfield  {journal} {\bibinfo
   {journal} {Phys. Rev. B}\ }\textbf {\bibinfo {volume} {101}},\ \bibinfo
  {pages} {224411} (\bibinfo {year} {2020})}\BibitemShut {NoStop}%
\end{thebibliography}

%

\end{document}